\documentclass[12pt] {article}
\usepackage{psfrag}
\usepackage{graphicx}
\usepackage{latexsym,amsfonts}
\usepackage{amssymb}
\begin{document}
\setcounter{page}{1}
\def\theequation{\arabic{section}.\arabic{equation}}
\def\theequation{\thesection.\arabic{equation}}
\setcounter{section}{0}

\title{On the renormalization of the sine--Gordon model}

\author{H. Bozkaya\,\thanks{E--mail: hidir@kph.tuwien.ac.at, Tel.:
+43--1--58801--14262, Fax: +43--1--58801--14299}\,, M. Faber\,\thanks{E--mail: 
faber@kph.tuwien.ac.at, Tel.:
+43--1--58801--14261, Fax: +43--1--58801--14299}\,,
A. N. Ivanov\,\thanks{E--mail: ivanov@kph.tuwien.ac.at, Tel.:
+43--1--58801--14261, Fax: +43--1--58801--14299}~\thanks{Permanent
Address: State Polytechnic University, Department of Nuclear
Physics, 195251 St. Petersburg, Russian Federation}\,, 
M. Pitschmann\,\thanks{E--mail: pitschmann@kph.tuwien.ac.at, Tel.:
+43--1--58801--14263, Fax: +43--1--58801--14299}}

\date{\today}

\maketitle
\vspace{-0.5in}
\begin{center}
{\it Atominstitut der \"Osterreichischen Universit\"aten,
Arbeitsbereich Kernphysik und Nukleare Astrophysik, Technische
Universit\"at Wien, \\ Wiedner Hauptstrasse 8-10/142, A-1040 Wien,
\"Osterreich }
\end{center}

\begin{center}
\begin{abstract}
  We analyse the renormalizability of the sine--Gordon model by the
  example of the two--point causal Green function up to second order
  in $\alpha_r(M^2)$, the dimensional coupling constant defined at the
  normalization scale $M$, and to all orders in $\beta^2$, the
  dimensionless coupling constant. We show that all divergences can be
  removed by the renormalization of the dimensional coupling constant
  using the renormalization constant $Z_1$, calculated in (J. Phys. A
  {\bf 36}, 7839 (2003)) within the path--integral approach. We show
  that after renormalization of the two--point Green function to first
  order in $\alpha_r(M^2)$ and to all orders in $\beta^2$ all higher
  order corrections in $\alpha_r(M^2)$ and arbitrary orders in
  $\beta^2$ can be expressed in terms of $\alpha_{\rm ph}$, the
  physical dimensional coupling constant independent on the
  normalization scale $M$. We calculate the Gell--Mann--Low function
  and analyse the dependence of the two--point Green function on
  $\alpha_{\rm ph}$ and the running coupling constant within the
  Callan--Symanzik equation. We analyse the renormalizability of
  Gaussian fluctuations around a soliton solution.  We show that
  Gaussian fluctuations around a soliton solution are renormalized
  like quantum fluctuations around the trivial vacuum to first orders
  in $\alpha_r(M^2)$ and $\beta^2$ and do not introduce any
  singularity to the sine--Gordon model at $\beta^2 = 8\pi$. The
  finite correction to the soliton mass, coinciding with that
  calculated by Dashen {\it et al.} (Phys.  Rev.  D {\bf 10}, 4130
  (1974)), appears in our approach to second order in $\alpha_{\rm
    ph}$ and to first order in $\beta^2$.  This is a perturbative
  correction, which provides no singularity for the sine--Gordon model
  at $\beta^2 = 8\pi$.  We calculate the correction to the soliton
  mass, caused by Gaussian fluctuations around a soliton, within the
  discretization procedure for various boundary conditions and find
  complete agreement with our result, obtained in continuous
  space--time.
\end{abstract}
\end{center}

\newpage

\section{Introduction}
\setcounter{equation}{0}

The sine--Gordon model we describe by the Lagrangian \cite{SC75,FI6}
\begin{eqnarray}\label{label1.1}
{\cal L}(x) = \frac{1}{2}\,\partial_{\mu}\vartheta(x)
\partial^{\mu}\vartheta(x) +
\frac{\alpha_0(\Lambda^2)}{\beta^2}\,(\cos\beta\vartheta(x) - 1),
\end{eqnarray}
where the field $\vartheta(x)$ and the coupling constant $\beta$ are
unrenormalizable, $\alpha_0(\Lambda^2)$ is a dimensional {\it bare}
coupling constant and $\Lambda$ is an ultra--violet cut--off. As has
been shown in \cite{FI6} the coupling constant $\alpha_0(\Lambda^2)$
is multiplicatively renormalizable and the renormalized Lagrangian
reads \cite{FI6}
\begin{eqnarray}\label{label1.2}
{\cal L}(x) &=& \frac{1}{2}\partial_{\mu}\vartheta(x)
\partial^{\mu}\vartheta(x) +
\frac{\alpha_r(M^2)}{\beta^2}(\cos\beta\vartheta(x) - 1) + (Z_1 -
1)\frac{\alpha_r(M^2)}{\beta^2}(\cos\beta\vartheta(x) - 1) =
\nonumber\\ &=& \frac{1}{2}\,\partial_{\mu}\vartheta(x)
\partial^{\mu}\vartheta(x) +
Z_1\,\frac{\alpha_r(M^2)}{\beta^2}\,(\cos\beta\vartheta(x) - 1),
\end{eqnarray}
where $Z_1 = Z_1(\alpha_r(M^2),\beta^2, M^2; \Lambda^2)$ is the
renormalization constant \cite{FI6}--\cite{SW96} depending on the
normalization scale $M$. The renormalization constant relates the
renormalized coupling constant $\alpha_r(M^2)$, depending on the
normalization scale $M$, to the {\it bare} coupling constant
$\alpha_0(\Lambda^2)$ \cite{FI6}--\cite{SW96}:
\begin{eqnarray}\label{label1.3}
  \alpha_r(M^2) = Z^{-1}_1(\alpha_r(M^2),\beta^2, M^2;\Lambda^2)\,
\alpha_0(\Lambda^2).
\end{eqnarray}
As has been found in \cite{FI6} the renormalization constant
$Z_1(\alpha_r(M^2), \beta^2, M^2; \Lambda)$ is equal to
\begin{eqnarray}\label{label1.4}
Z_1(\alpha_r(M^2), \beta^2, M^2; \Lambda^2) =
\Big(\frac{\Lambda^2}{M^2}\Big)^{\beta^2/8\pi}.
\end{eqnarray}
This result is valid to all orders of perturbation theory developed
relative to the coupling constant $\beta^2$ and $\alpha_0(\Lambda^2)$
\cite{FI6}.  Since the normalization constant does not depend on
$\alpha_r(M^2)$, we write below $Z_1 = Z_1(\beta^2,M^2; \Lambda^2)$.

For the analysis of the renormalizability of the sine--Gordon model
with respect to quantum fluctuations around the trivial vacuum we
expand the Lagrangian (\ref{label1.2}) in powers of $\vartheta(x)$.
This gives
\begin{eqnarray}\label{label1.5}
{\cal L}(x) &=& \frac{1}{2}\,[\partial_{\mu}\vartheta(x)
\partial^{\mu}\vartheta(x) - \alpha_r(M^2)\,\vartheta^2(x)] 
+ {\cal L}_{\rm int}(x),
\end{eqnarray}
where ${\cal L}_{\rm int}(x)$ describes the self--interactions of the
sine--Gordon field
\begin{eqnarray}\label{label1.6}
{\cal L}_{\rm int}(x) &=&  \alpha_r(M^2)\sum^{\infty}_{n = 2
}\frac{(-1)^n}{(2n)!}\beta^{2(n -
  1)}\vartheta^{2n}(x)\nonumber\\ 
&+&  (Z_1 -
1)\,\alpha_r(M^2)\sum^{\infty}_{n = 1}\frac{(-1)^n}{(2n)!}
\beta^{2(n -
  1)}\vartheta^{2n}(x).
\end{eqnarray}
It is seen that the coupling constant $\alpha_r(M^2)$ has the meaning of
a squared mass of free quanta of the sine--Gordon field
$\vartheta(x)$.  The causal two--point Green function of free
sine--Gordon quanta with mass $\alpha_r(M^2)$ is defined by
\begin{eqnarray}\label{label1.7}
  -i\,\Delta_F(x; \alpha_r(M^2)) = 
\langle 0|{\rm T}(\vartheta(x)\vartheta(0))|0\rangle = 
  \int \frac{d^2k}{(2\pi)^2 i}\,\frac{\displaystyle
 e^{\textstyle -i\,k\cdot x}}{\alpha_r(M^2) - k^2 -i\,0}.
\end{eqnarray}
At $x = 0$ the Green function $-i\,\Delta_F(0;
\alpha_r(M^2))$ is equal to \cite{FI6}
\begin{eqnarray}\label{label1.8}
  -i\,\Delta_F(0; \alpha_r(M^2)) = \frac{1}{4\pi}\,{\ell n}
\Big[\frac{\Lambda^2}{\alpha_r(M^2)}\Big],
\end{eqnarray}
where $\Lambda$ is a cut--off in Euclidean 2--dimensional momentum
space \cite{FI6}.

The paper is organized as follows. In Section 2 we analyse the
renormalizability of the sine--Gordon model by means of power
counting.  In Sections 3 we investigate the renormalizability of the
two--point Green function of the sine--Gordon field. Quantum
fluctuations are calculated relative to the trivial vacuum up to
second order in $\alpha_r(M^2)$ and to all orders in $\beta^2$. We
show that after renormalization of the two--point Green function to
first order in $\alpha_r(M^2)$ and to all orders in $\beta^2$ all
higher order corrections in $\alpha_r(M^2)$ and all orders in
$\beta^2$ can be expressed in terms of $\alpha_{\rm ph}$, the physical
dimensional coupling constant independent on the normalization scale
$M$. We derive the effective Lagrangian of the sine--Gordon model,
taking into account quantum fluctuations to second order in
$\alpha_r(M^2)$ and to all orders in $\beta^2$. We show that the
correction of second order in $\alpha_{\rm ph}$ and first order in
$\beta^2$, i.e.  $O(\alpha^2_{\rm ph}\beta^2)$, reproduces the finite
contribution $-\,\sqrt{\alpha_{\rm ph}}/\pi$ to the soliton mass
coinciding with that calculated by Dashen {\it et al.}
\cite{RD74,RD75}.  In our approach this is a perturbative correction,
which does not lead to a singularity of the sine--Gordon model at
$\beta^2 = 8\pi$.  This confirms the absence of a singularity of the
renormalized sine--Gordon model at $\beta^2 = 8\pi$ conjectured by
Zamolodchikov and Zamolodchikov \cite{AZ79} and proved in \cite{FI6}.
In Section 4 we analyse the renormalizability of the sine--Gordon
model within the Renormalization Group approach.  We use the
Callan--Symanzik equation for the derivation of the total two--point
Green function of the sine--Gordon field in the momentum
representation.  We show that the two--point Green function depends on
the running coupling constant $\alpha_r(p^2) = \alpha_{\rm
  ph}(p^2/\alpha_{\rm ph})^{\tilde{\beta}^2/8\pi}$, where
$\tilde{\beta}^2 = \beta^2/(1 + \beta^2/8\pi) < 1$ for all $\beta^2$.
In Section 5 we investigate the renormalizability of the sine--Gordon
model with respect to Gaussian fluctuations around a soliton solution.
We show that Gaussian fluctuations around a soliton solution lead to
the same renormalized Lagrangian of the sine--Gordon model as quantum
fluctuations around the trivial vacuum taken into account to first
order in $\alpha_r(M^2)$ and $\beta^2$.  In Section 6 we discuss the
correction to the soliton mass induced by quantum fluctuations.  We
show that Gaussian fluctuations around a soliton solution reproduce
the same correction as the quantum fluctuations around the trivial
vacuum, calculated to first orders in $\alpha_r(M^2)$ and $\beta^2$.
This correction does not contain the non--perturbative finite quantum
correction obtained by Dashen {\it et al.}  \cite{RD74,RD75}. In our
analysis of the sine--Gordon model such a finite correction appears
only to second order in $\alpha_{\rm ph}$ and to first order in
$\beta^2$ (see Section 3). In Section 7 we discuss the calculation of
the correction to the soliton mass $\Delta M_s$, induced by Gaussian
fluctuations, within a discretization procedure for various boundary
conditions. We show that the result of the calculation of $\Delta M_s$
does not depend on the boundary conditions and agrees fully with that
obtained in continuous space--time.  In the Conclusion we summarize
the obtained results and discuss them.  In the Appendix we adduce the
solutions of the differential equation related to the calculation of
Gaussian fluctuations around a soliton.

\section{Power counting and renormalization of the sine--Gordon 
model }
\setcounter{equation}{0}

As usual the general analysis of renormalizability of a quantum field
theory is carried out in the form of {\it power counting}, the concept
of the superficial degree of divergence of momentum integrals based on
dimensional considerations \cite{IZ80}--\cite{SW96}.

The analysis of the convergence of a given Feynman diagram $G$ within
{\it power counting} is done by scaling all internal momenta with a
common factor $\lambda$, $k_{\ell} \to \lambda k_{\ell}$, and looking
at the behaviour $I_G \sim \lambda^{\omega(G)}$ at $\lambda \to
\infty$.  Since we deal with a quantum field theory of a
(pseudo)scalar field, the propagator of such a field behaves as
$\lambda^{-2}$ at $\lambda \to \infty$.

Let a given Feynman diagram $G$ contain $L$ independent loops, $I$
internal boson lines and $V_{2n}$ vertices with $2n$ lines. Since we
have no vertices with derivatives of the sine--Gordon field, the
superficial degree of divergence $\omega(G)$ of a diagram $G$ is
\cite{IZ80}--\cite{SW96}
\begin{eqnarray}\label{label2.1}
\omega(G) = 2L - 2 I.
\end{eqnarray}
The number of independent loops $L$ is defined by \cite{IZ80}--\cite{SW96}
\begin{eqnarray}\label{label2.2}
L = I + 1 - \sum_{\{n\}}V_{2n},
\end{eqnarray}
where the sum extends over all vertices defining the Feynman diagram
$G$.  Substituting (\ref{label2.2}) into (\ref{label2.1}) we can
express the superficial degree of divergence $\omega(G)$ in terms of
vertices only
\begin{eqnarray}\label{label2.3}
\omega(G) = 2 - 2\sum_{\{n\}}V_{2n}.
\end{eqnarray}
This testifies the complete renormalizability of the sine--Gordon
model.  

Indeed, the vacuum energy density is quadratically divergent, since it
corresponds to the ``Feynman diagram without vertices''.  Such a
quadratic ultra--violet divergence of the vacuum energy density of the
sine--Gordon model has been recently shown in \cite{FI6}. Such a
quadratic divergence of the vacuum energy density can be removed by
normal--ordering the operator of the Hamilton density of the
sine--Gordon model \cite{FI6}. 

All Feynman diagrams with one vertex diverge logarithmically, and any
Feynman diagram with more than one vertex converges.  As has been
shown in \cite{FI6} by using the path--integral approach, all
logarithmic divergences can be removed by the renormalization of the
dimensional coupling constant $\alpha_0(\Lambda^2)$.

\section{Renormalization of  causal two--point Green function}
\setcounter{equation}{0}

The causal two--point Green function of the sine--Gordon field is defined
by
\begin{eqnarray}\label{label3.1}
-i\,\Delta(x) = \frac{1}{i}\frac{\delta}{\delta
J(x)}\frac{1}{i}\frac{\delta}{\delta J(0)}Z[J]_{J = 0},
\end{eqnarray}
where $Z[J]$ is a generating functional of Green functions
\begin{eqnarray}\label{label3.2}
  \hspace{-0.3in}&&Z[J] = \int {\cal D}\vartheta\,
  \exp\Big\{\,i\int d^2y\,[{\cal L}(y)
  + \vartheta(y)J(y)]\Big\} = \nonumber\\
  \hspace{-0.3in}&&=\int {\cal D}\vartheta\,
  \exp\Big\{\,i\int d^2y\,\Big[\frac{1}{2}\,\Big(\partial_{\mu}\vartheta(y)
  \partial^{\mu}\vartheta(y) - \alpha_r(M^2)\,\vartheta^2(y)\Big) + 
  {\cal L}_{\rm int}(y) + \vartheta(y)J(y)\Big]\Big\},\quad\quad
\end{eqnarray}
normalized by $Z[0] = 1$, $J(x)$ is the external source of the
sine--Gordon field $\vartheta(x)$. 

Substituting (\ref{label3.2}) into (\ref{label3.1}) we get
\begin{eqnarray}\label{label3.3}
  -i\,\Delta(x) &=& \int {\cal D}\vartheta\,\vartheta(x)
  \vartheta(0)\,\exp\Big\{\,i\int d^2y\,{\cal L}_{\rm int}(y)\Big\} 
  \nonumber\\
  &\times&
  \exp\Big\{\,\frac{i}{2}\int d^2y\,\Big[\partial_{\mu}\vartheta(y)
  \partial^{\mu}\vartheta(y) - \alpha_r(M^2)\,\vartheta^2(y) \Big]\Big\},
\end{eqnarray}
where ${\cal L}_{\rm int}(y)$ is given by (\ref{label1.6}).  The
r.h.s. of (\ref{label3.3}) can be rewritten in the form of a vacuum
expectation value of a time--ordered product
\begin{eqnarray}\label{label3.4}
  -i\,\Delta(x) = \langle 0|{\rm T}\Big(\vartheta(x)\vartheta(0)\,
\exp\Big\{\,i\int d^2y\,:{\cal L}_{\rm int}(y):\Big\}\Big)
|0\rangle_{c},   
\end{eqnarray}
where the index $c$ means the {\it connected} part, $:\ldots:$ denotes
normal ordering, $\vartheta(x)$ is the free sine--Gordon field
operator with mass $\alpha_r(M^2)$ and the causal two--point Green
function $\Delta_F(x,\alpha_r(M^2))$  defined by (\ref{label1.7}).

In the momentum representation the two--point Green function
(\ref{label3.4}) reads
\begin{eqnarray}\label{label3.5}
  -i\,\tilde{\Delta}(p) &=& -i\int d^2x\,e^{\textstyle +ip\cdot x}\Delta(x) = 
  \nonumber\\
  &=&\int d^2x\,e^{\textstyle +ip\cdot x}\,\langle 0|{\rm T}\Big(\vartheta(x)
  \vartheta(0)\,\exp\Big\{\,i\int d^2y\,:{\cal L}_{\rm int}(y):\Big\}\Big)
  |0\rangle_{c}.   
\end{eqnarray}
For the analysis of the renormalizability of the sine--Gordon model we
propose to calculate the corrections to the two--point Green function
(\ref{label3.4}) (or to (\ref{label3.5})), induced by quantum
fluctuations around the trivial vacuum.  Expanding the r.h.s.  of
Eq.(\ref{label3.4}) in powers of $\alpha_r(M^2)$ and $\beta^2$ we
determine
\begin{eqnarray}\label{label3.6}
  -i\,\Delta(x)= \sum^{\infty}_{m = 0}(- i)\,\Delta^{(m)}(x,\alpha_r(M^2)),
\end{eqnarray}
where $(- i)\,\Delta^{(m)}(x,\alpha_r(M^2))$ is defined by
\begin{eqnarray}\label{label3.7}
  -i\,\Delta^{(m)}(x,\alpha_r(M^2)) = \frac{i^m}{m!}\int
  \prod^{m}_{k = 1}d^2y_k\langle 0|{\rm T}(\vartheta(x)
  \vartheta(0):{\cal L}_{\rm int}(y_k):)|0\rangle_{c}.   
\end{eqnarray}
The Green function $(-i)\,\Delta^{(0)}(x,\alpha_r(M^2))$ coincides
with the Green function (\ref{label1.7}) of the free sine--Gordon
field.

In the momentum representation the correction to the two--point Green
function $(- i)\,\Delta^{(m)}(x,\alpha_r(M^2))$ can be written as
\begin{eqnarray}\label{label3.8}
 && -i\,\tilde{\Delta}^{(m)}(p,\alpha_r(M^2)) =  \int d^2x\,
e^{\textstyle +ip\cdot x}\,(-i)\,\Delta^{(m)}(x,\alpha_r(M^2)) = \nonumber\\
&&= \frac{i^m}{m!}\int d^2x\,e^{\textstyle +ip\cdot x}\int
  \prod^{m}_{k = 1}d^2y_k\langle 0|{\rm T}(\vartheta(x)
  \vartheta(0):{\cal L}_{\rm int}(y_k):)|0\rangle_{c}.   
\end{eqnarray}
The momentum representation is more convenient for the perturbative
analysis of the renormalization of the two--point Green function of
the sine--Gordon field.

\subsection{Two--point Green function to first order in
  $\alpha_r(M^2)$ and to all orders in $\beta^2$}

The correction to the two--point Green function to first order in
$\alpha_r(M^2)$ and to all orders in $\beta^2$ is defined by
\begin{eqnarray}\label{label3.9}
  \hspace{-0.3in}&&-i\,\Delta^{(1)}(x,\alpha_r(M^2)) = i\int
  d^2y_1\langle 0|{\rm T}(\vartheta(x)
  \vartheta(0):{\cal L}_{\rm int}(y_1):)|0\rangle_{c}=\nonumber\\
\hspace{-0.3in}&&=  i\,\alpha_r(M^2)\sum^{\infty}_{n = 2
  }\frac{(-1)^n}{(2n)!}\beta^{2(n -
    1)}\int d^2y\,\langle 0|{\rm T}(\vartheta(x)
  \vartheta(0):\vartheta^{2n}(y):)|0\rangle_{c}\nonumber\\
  \hspace{-0.3in}&&+  i\,\alpha_r(M^2)\,(Z_1 - 1)\sum^{\infty}_{n = 1
  }\frac{(-1)^n}{(2n)!}\beta^{2(n -
    1)}\int d^2y\,\langle 0|{\rm T}(\vartheta(x)
  \vartheta(0):\vartheta^{2n}(y):)|0\rangle_{c}.
\end{eqnarray}
Making all contractions we arrive at the expression
\begin{eqnarray}\label{label3.10}
  \hspace{-0.3in}&&-i\,\Delta^{(1)}(x,\alpha_r(M^2))  = i\,\alpha_r(M^2)\sum^{\infty}_{n = 2
  }\frac{(-1)^n}{(2n)!}\beta^{2(n -
    1)}[2n\,(2n-1)!!][-i\Delta_F(0,\alpha_r(M^2))]^{n-1
  }\nonumber\\ 
   \hspace{-0.3in} &&\times\int d^2y\,[-i\Delta_F(x - y,\alpha_r(M^2))]
  [-i\Delta_F(-y,\alpha_r(M^2))]
  \nonumber\\
   \hspace{-0.3in}&&+  
  i\,\alpha_r(M^2)\,(Z_1 - 1)\sum^{\infty}_{n = 1
  }\frac{(-1)^n}{(2n)!}\beta^{2(n -
    1)}[2n\,(2n-1)!!][-i\Delta_F(0,\alpha_r(M^2))]^{n-1
  }\nonumber\\ 
   \hspace{-0.3in} &&\times\int d^2y\,[-i\Delta_F(x - y,\alpha_r(M^2))
  ][-i\Delta_F(-y,\alpha_r(M^2))]
\end{eqnarray}
The sums over $n$ are equal to
\begin{eqnarray}\label{label3.11}
  \hspace{-0.3in}&&\sum^{\infty}_{n = 2}\frac{(-1)^n}{(2n)!}[2n\,(2n-1)!!]
  [\beta^2(-i)\Delta_F(0,\alpha_r(M^2))]^{n-1} =  1 - \exp\Big\{\frac{1}{2}\,
  \beta^2 i\Delta_F(0,\alpha_r(M^2))\Big\},\nonumber\\
  \hspace{-0.3in}&&\sum^{\infty}_{n = 1
  }\frac{(-1)^n}{(2n)!}\beta^{2(n -
    1)}[2n\,(2n-1)!!][-i\Delta_F(0,\alpha_r(M^2))]^{n-1
  } = - \exp\Big\{\frac{1}{2}\,
  \beta^2 i\Delta_F(0,\alpha_r(M^2))\Big\}.\nonumber\\
  \hspace{-0.3in}&&
\end{eqnarray}
Substituting (\ref{label3.11}) into (\ref{label3.10}) we get
\begin{eqnarray}\label{label3.12}
  -\,i\,\Delta^{(1)}(x,\alpha_r(M^2)) &=&i\,\alpha_r(M^2)\,
\Big[1 - \exp\Big\{\frac{1}{2}\,
  \beta^2 i \Delta_F(0,\alpha_r(M^2))\Big\}\Big]\nonumber\\ 
  &\times& \int d^2y\,[-i\Delta_F(x - y,\alpha_r(M^2))]
  [-i\Delta_F(-y,\alpha_r(M^2))]
  \nonumber\\
  &-& 
  i\,\alpha_r(M^2)\,(Z_1 - 1)\,\exp\Big\{\frac{1}{2}\,
  \beta^2 i \Delta_F(0,\alpha_r(M^2))\Big\}\nonumber\\
  &\times&\int d^2y\,[-i\Delta_F(x - y,\alpha_r(M^2))
  ][-i\Delta_F(-y,\alpha_r(M^2))].
\end{eqnarray}
The r.h.s. of (\ref{label3.12}) can be transcribed into the form
\begin{eqnarray}\label{label3.13}
  -\,i\,\Delta^{(1)}(x,\alpha_r(M^2)) &=& i\,\alpha_r(M^2)\,
  \Big[1 - Z_1\,\exp\Big\{\frac{1}{2}\,
  \beta^2 i \Delta_F(0,\alpha_r(M^2))\Big\}\Big]\nonumber\\
  &\times&\int d^2y\,[-i\Delta_F(x - y,\alpha_r(M^2))]
  [-i\Delta_F(-y,\alpha_r(M^2))].
  \end{eqnarray}
  Using the normalization constant $Z_1$, given by (\ref{label1.4}),
  and the definition (\ref{label1.8}) of the two--point Green function
  we remove the cut--off $\Lambda$
\begin{eqnarray}\label{label3.14}
  -\,i\,\Delta^{(1)}(x,\alpha_r(M^2)) &=&  i\,\alpha_r(M^2)\,
  \Big[1 - \Big(\frac{\alpha_r(M^2)}{M^2}\Big)^{\beta^2/8\pi}\Big]\nonumber\\
  &\times&\int d^2y\,[-i\Delta_F(x - y,\alpha_r(M^2))]
  [-i\Delta_F(-y,\alpha_r(M^2))].
  \end{eqnarray}
  Thus, the renormalized causal two--point Green function of the
  sine--Gordon field, defined to first order in $\alpha_r(M^2)$ and to
  all orders in $\beta^2$ is given by
\begin{eqnarray}\label{label3.15}
  -\,i\,\Delta(x) &=& -\,i\,\Delta_F(x,\alpha_r(M^2)) +  i\,\alpha_r(M^2)\,
  \Big[1 - \Big(\frac{\alpha_r(M^2)}{M^2}\Big)^{\beta^2/8\pi}\Big]\nonumber\\
  &\times&\int d^2y\,[-i\Delta_F(x - y,\alpha_r(M^2))]
  [-i\Delta_F(-y,\alpha_r(M^2))].
  \end{eqnarray}
  In the momentum representation the two--point Green function
  (\ref{label3.15}) reads
\begin{eqnarray}\label{label3.16}
  \hspace{-0.3in}&&-\, i\, \tilde{\Delta}(p) = 
  \frac{(-i)}{\alpha_r(M^2) - p^2} +  
  \frac{(-i)}{\alpha_r(M^2) - p^2}\, i\, \alpha_r(M^2)\,
  \Big[1 - \Big(\frac{\alpha_r(M^2)}{M^2}\Big)^{\beta^2/8\pi}\Big]
  \frac{(-i)}{\alpha_r(M^2) - p^2}.\nonumber\\
  \hspace{-0.3in}&&
  \end{eqnarray}
The second term defines the correction to the mass of the sine--Gordon field
\begin{eqnarray}\label{label3.17}
\delta \alpha_r(M^2) = -\,\alpha_r(M^2)\,
  \Big[1 - \Big(\frac{\alpha_r(M^2)}{M^2}\Big)^{\beta^2/8\pi}\Big].
  \end{eqnarray}
  Thus, the two--point Green function, calculated to first order in
  $\alpha_r(M^2)$ and to all orders in $\beta^2$ is equal to
\begin{eqnarray}\label{label3.18}
  \hspace{-0.3in}-\,i\,\tilde{\Delta}(p)= \frac{(-i)}{\alpha_r(M^2) + 
\delta \alpha_r(M^2)  - p^2} = \frac{(-i)}{\alpha_{\rm ph}  - p^2},
  \end{eqnarray}
  where $\alpha_{\rm ph}$ is determined by
\begin{eqnarray}\label{label3.19}
  \alpha_{\rm ph} = \alpha_r(M^2) + \delta \alpha_r(M^2) = \alpha_r(M^2)\,
\Big(\frac{\alpha_r(M^2)}{M^2}\Big)^{\beta^2/8\pi}.
  \end{eqnarray}
This gives  also $\alpha_r(M^2)$ in term of $M$ and $\alpha_{\rm ph}$:
\begin{eqnarray}\label{label3.20}
 \alpha_r(M^2) = \alpha_{\rm ph} 
\Big(\frac{M^2}{\alpha_{\rm ph}}\Big)^{\tilde{\beta}^2/8\pi}\quad,\quad \tilde{\beta}^2 = \frac{\beta^2}{\displaystyle 1 + \frac{\beta^2}{8\pi}}.
  \end{eqnarray}
  The Green function (\ref{label3.18}) can be obtained to leading
  order in $\beta^2$ from the Lagrangian
\begin{eqnarray}\label{label3.21}
  {\cal L}_{\rm eff}(x) &=& \frac{1}{2}\,\partial_{\mu}\vartheta(x)
  \partial^{\mu}\vartheta(x) + 
  \frac{\alpha_r(M^2)}{\beta^2}\Big(\frac{\alpha_r(M^2)}{M^2}\Big)^{\beta^2/8\pi}
  \,(\cos\beta\vartheta(x) - 1) = \nonumber\\
  &=&\frac{1}{2}\,\partial_{\mu}\vartheta(x)
  \partial^{\mu}\vartheta(x) + 
  \frac{\alpha_{\rm ph}}{\beta^2}
  \,(\cos\beta\vartheta(x) - 1). 
  \end{eqnarray}
  We argue that higher order corrections to the two--point Green
  function in $\alpha_r(M^2)$ and to all orders in $\beta^2$ should
  depend on the physical coupling constant $\alpha_{\rm ph}$ only
\begin{eqnarray}\label{label3.22}
  -i\,\Delta^{(m)}(x,\alpha_r(M^2)) &=& \,\frac{i^m}{m!}\int
  \prod^{m}_{k = 1}d^2y_k\langle 0|{\rm T}(\vartheta(x)
  \vartheta(0):{\cal L}_{\rm int}(y_k):)|0\rangle_{c} =\nonumber\\
  &=&  -i\,\Delta^{(m)}(x,
  \alpha_{\rm ph})\quad ({\rm for} \quad m \ge 2).  
\end{eqnarray}
In order to prove this assertion it is sufficient to analyse the
renormalization of the causal two--point Green function to second
order in $\alpha_r(M^2)$ and to all orders in $\beta^2$.

  \subsection{Two--point Green function to second order in
    $\alpha_r(M^2)$ and to all orders in $\beta^2$}

  The correction to the two--point Green function to second order in
  $\alpha_r(M^2)$ and to all orders in $\beta^2$ is defined by
\begin{eqnarray}\label{label3.23}
  -i\,\Delta^{(2)}(x,\alpha_r(M^2)) &=&  -\,\frac{1}{2}\int\!\!\!\int 
d^2y_1d^2y_2\,
  \langle 0|{\rm T}(\vartheta(x)
  \vartheta(0):{\cal L}_{\rm int}(y_1)::{\cal L}_{\rm int}(y_2):)|0\rangle_{c}=
\nonumber\\
&=& -\,\frac{1}{2}\,\alpha^2_r(M^2)\sum^{\infty}_{n_1 = 2
  }\frac{(-1)^{n_1}}{(2n_1)!}\beta^{2(n_1 -
    1)}\sum^{\infty}_{n_2 = 2
  }\frac{(-1)^{n_2}}{(2n_2)!}\beta^{2(n_2 -
    1)}\nonumber\\
&\times&\int\!\!\!\int d^2y_1d^2y_2\,
  \langle 0|{\rm T}(\vartheta(x)
  \vartheta(0):\vartheta^{2n_1}(y_1)::\vartheta^{2n_2}(y_2):)|0\rangle_{c}\nonumber\\
&-&\alpha^2_r(M^2)(Z_1 - 1)\sum^{\infty}_{n_1 = 2
  }\frac{(-1)^{n_1}}{(2n_1)!}\beta^{2(n_1 -
    1)}\sum^{\infty}_{n_2 = 1
  }\frac{(-1)^{n_2}}{(2n_2)!}\beta^{2(n_2 -
    1)}\nonumber\\
&\times&\int\!\!\!\int d^2y_1d^2y_2\,
  \langle 0|{\rm T}(\vartheta(x)
  \vartheta(0):\vartheta^{2n_1}(y_1)::\vartheta^{2n_2}(y_2):)|0\rangle_{c}\nonumber\\
&-&\frac{1}{2}\,\alpha^2_r(M^2)(Z_1 - 1)^2\sum^{\infty}_{n_1 = 1
  }\frac{(-1)^{n_1}}{(2n_1)!}\beta^{2(n_1 -
    1)}\sum^{\infty}_{n_2 = 1
  }\frac{(-1)^{n_2}}{(2n_2)!}\beta^{2(n_2 -
    1)}\nonumber\\
&\times&\int\!\!\!\int d^2y_1d^2y_2\,
  \langle 0|{\rm T}(\vartheta(x)
  \vartheta(0):\vartheta^{2n_1}(y_1)::\vartheta^{2n_2}(y_2):)|0\rangle_{c}.
\end{eqnarray}
This expression can be described in terms of two classes of
topologically different Feynman diagrams depicted in Fig.1.
\begin{figure}[h]
 \centering
 \psfrag{p}{$p$}
 \psfrag{0}{$x$}
 \psfrag{x}{$0$}
  \psfrag{y1}{$y_1$}
  \psfrag{y2}{$y_2$}
  \psfrag{q1}{$q_1$}
  \psfrag{qi}{$q_i$}
  \psfrag{qi+1}{$q_{i+1}$}
  \psfrag{qi+2}{}
  \psfrag{qi+3}{}
  \psfrag{qj-1}{}
  \psfrag{qj-2}{}
  \psfrag{qj}{$q_{j}$}
  \psfrag{qj+1}{$q_{j+1}$}
  \psfrag{q2n}{$q_{2n}$}
   \includegraphics[scale=0.25]{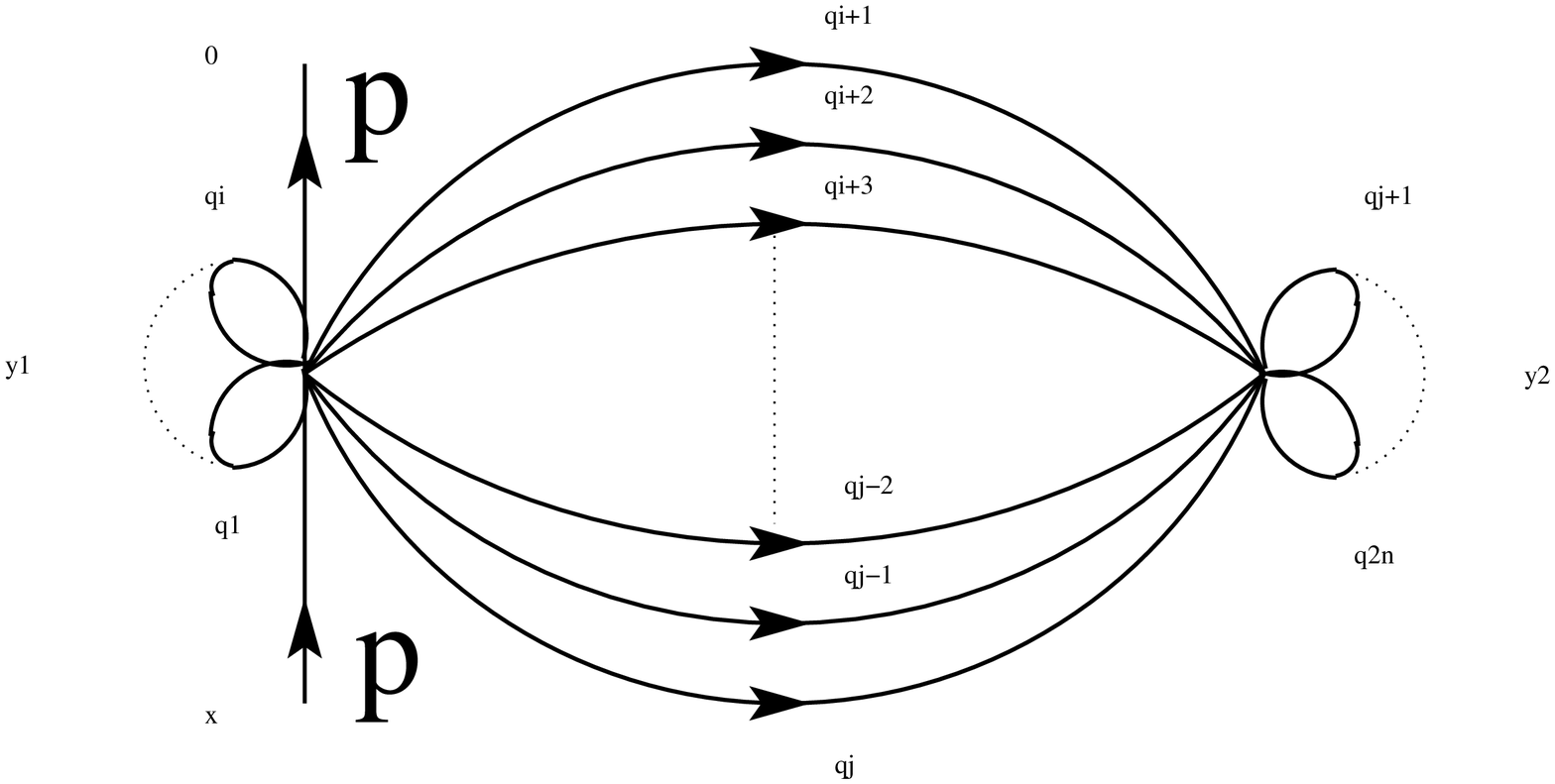} \hspace{1.5cm}
   \includegraphics[scale=0.25]{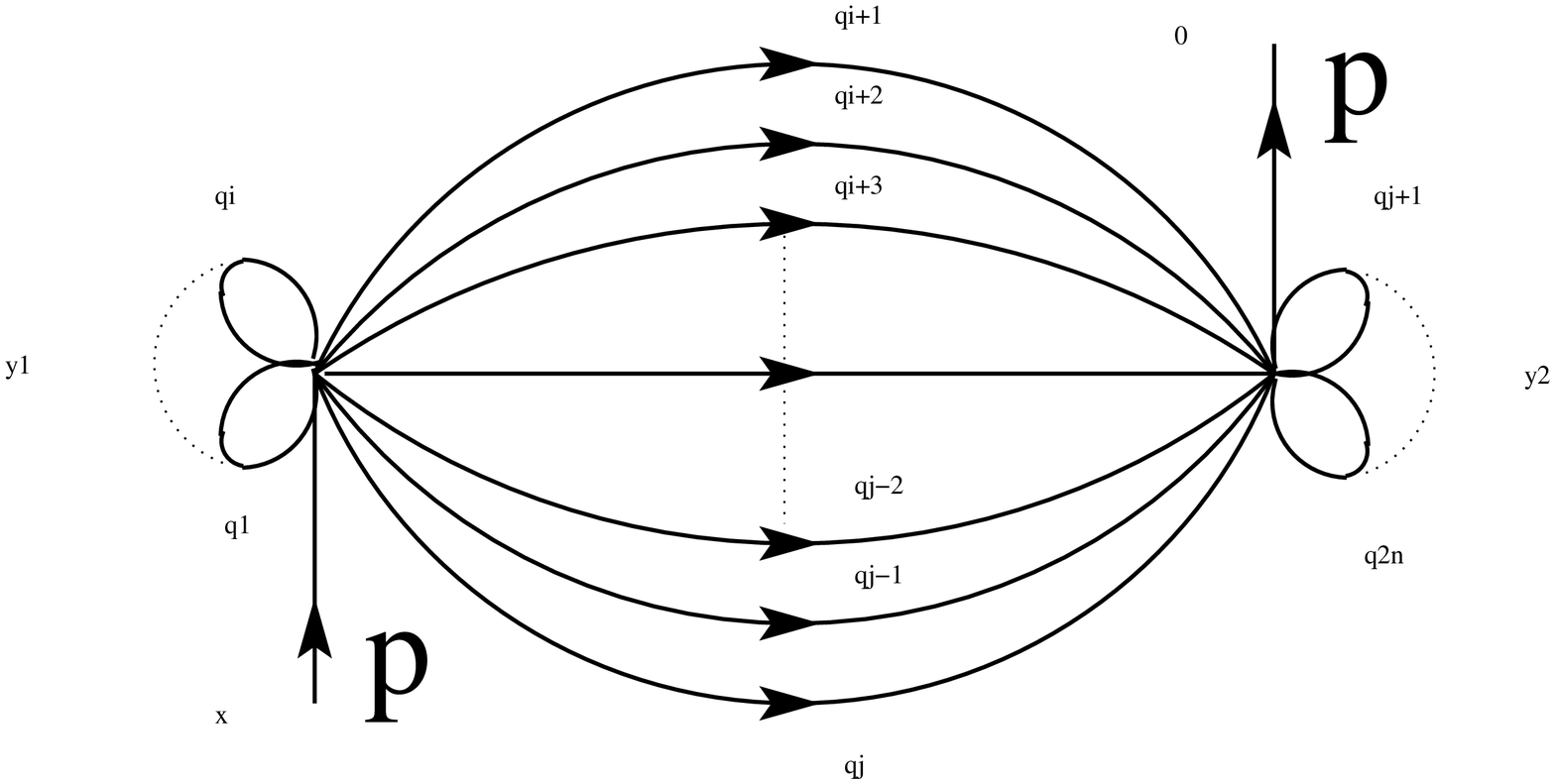}
   \caption{Feynman diagrams for corrections to the two--point Green
     function to second order in $\alpha$ and to arbitrary order in
     $\beta^2$.  The left diagrams correspond to a non--vanishing
     expectation value for an even number of internal lines between
     the two vertices only, while the right diagrams describe a
     non--vanishing contribution for an odd number of internal lines
     between two vertices.}
\end{figure}
In the momentum representation this correction is equal to
\begin{eqnarray}\label{label3.24}
  && -i\,\tilde{\Delta}^{(2)}(p,\alpha_r(M^2)) = 
  i\,\Big[\alpha_r(M^2) Z_1 \exp\Big\{\frac{1}{2}\,
  \beta^2 i\Delta_F(0,\alpha_r(M^2))
  \Big\}\Big]^2\,\Big[\frac{(-i)}{\alpha_r(M^2) - p^2}\Big]^2\nonumber\\
  &&\times \Big\{\sum^{\infty}_{n = 1}\beta^{4n}
  \int \frac{d^2q_1}{(2\pi)^2 i}\,\frac{1}{\alpha_r(M^2) - q^2_1}
  \int \frac{d^2q_2}{(2\pi)^2 i}\,\frac{1}{\alpha_r(M^2) - q^2_2}\ldots \nonumber\\
  &&\times 
  \int \frac{d^2q_{2n}}{(2\pi)^2 i}\,\frac{1}{\alpha_r(M^2) - q^2_{2n}}
  \frac{1}{\alpha_r(M^2) - (p - q_1 - q_2 - \ldots - q_{2n})^2}\nonumber\\
  &&+\sum^{\infty}_{n = 0}\beta^{4n + 2}\int \frac{d^2q_1}{(2\pi)^2 i}\,\frac{1}{\alpha_r(M^2) - q^2_1}
  \int \frac{d^2q_2}{(2\pi)^2 i}\,\frac{1}{\alpha_r(M^2) - q^2_2}\ldots \nonumber\\
  &&\times 
  \int \frac{d^2q_{2n + 1}}{(2\pi)^2 i}\,\frac{1}{\alpha_r(M^2) - q^2_{2n + 1}}
  \frac{1}{\alpha_r(M^2) - (q_1 + q_2 + \ldots + q_{2n + 1})^2}\Big\} = \nonumber\\
  &&= 
  i\,\Big[\alpha_r(M^2)\Big(\frac{\alpha_r(M^2)}{M^2}\Big)^{\beta^2/8\pi}\Big]^2\,
  \Big[\frac{(-i)}{\alpha_r(M^2) - p^2}\Big]^2\nonumber\\
  &&\times \Big\{\sum^{\infty}_{n = 1}\beta^{4n}
  \int \frac{d^2q_1}{(2\pi)^2 i}\,\frac{1}{\alpha_r(M^2) - q^2_1}
  \int \frac{d^2q_2}{(2\pi)^2 i}\,\frac{1}{\alpha_r(M^2) - q^2_2}\ldots \nonumber\\
  &&\times 
  \int \frac{d^2q_{2n}}{(2\pi)^2 i}\,\frac{1}{\alpha_r(M^2) - q^2_{2n}}
  \frac{1}{\alpha_r(M^2) - (p - q_1 - q_2 - \ldots - q_{2n})^2}\nonumber\\
  &&+\sum^{\infty}_{n = 0}\beta^{4n + 2}\int \frac{d^2q_1}{(2\pi)^2 i}\,\frac{1}{\alpha_r(M^2) - q^2_1}
  \int \frac{d^2q_2}{(2\pi)^2 i}\,\frac{1}{\alpha_r(M^2) - q^2_2}\ldots \nonumber\\
  &&\times 
  \int \frac{d^2q_{2n + 1}}{(2\pi)^2 i}\,\frac{1}{\alpha_r(M^2) - q^2_{2n + 1}}
  \frac{1}{\alpha_r(M^2) - (q_1 + q_2 + \ldots + q_{2n + 1})^2}\Big\}. 
\end{eqnarray}
The common factor $[\alpha_r(M^2) Z_1 \exp\{ \beta^2
i\Delta_F(0,\alpha_r(M^2)/2)\}]^2$ is caused by the summation of the
infinite series of one--vertex--loop diagrams. Using (\ref{label1.4})
and (\ref{label3.19}) one can show that it is equal to $\alpha^2_{\rm
  ph}$:
\[\Big[\alpha_r(M^2) Z_1 \exp\Big\{ \frac{1}{2}\,\beta^2\,
i\Delta_F(0,\alpha_r(M^2))\Big\}\Big]^2 = \Big[\alpha_r(M^2)\Big(\frac{\alpha_r(M^2)}{M^2}\Big)^{\beta^2/8\pi}\Big]^2 = \alpha^2_{\rm ph}.\]
Thus, we have shown that after renormalization the correction to the
two--point Green function to second order in $\alpha_r(M^2)$ and to
all orders in the $\beta^2$ is proportional to $\alpha^2_{\rm ph}$.
Then, since the two--point Green function, calculated to first in
$\alpha_r(M^2)$ and to all orders in $\beta^2$, is given by
(\ref{label3.18}), in the correction of the second order
$\alpha_r(M^2)$ we can replace the two--point Green functions of the
free sine--Gordon fields in the r.h.s. of (\ref{label3.24}) by
(\ref{label3.18}). This gives
\begin{eqnarray}\label{label3.25}
  \hspace{-0.3in}&& -i\,\tilde{\Delta}^{(2)}(p,\alpha_r(M^2)) =
 i\,\alpha^2_{\rm ph}\,
  \Big[\frac{(-i)}{\alpha_{\rm ph} - p^2}\Big]^2\Big\{\sum^{\infty}_{n = 1}
\beta^{4n}
  \int \frac{d^2q_1}{(2\pi)^2 i}\,\frac{1}{\alpha_{\rm ph} - q^2_1}
  \int \frac{d^2q_2}{(2\pi)^2 i}\,\frac{1}{\alpha_{\rm ph} - q^2_2}
  \ldots \nonumber\\
  \hspace{-0.3in}&&\times 
  \int \frac{d^2q_{2n}}{(2\pi)^2 i}\,\frac{1}{\alpha_{\rm ph} - q^2_{2n}}
  \frac{1}{\alpha_{\rm ph} - (p - q_1 - q_2 - \ldots - q_{2n})^2}\nonumber\\
  \hspace{-0.3in}&&+\sum^{\infty}_{n = 0}\beta^{4n + 2}
  \int \frac{d^2q_1}{(2\pi)^2 i}
  \,\frac{1}{\alpha_{\rm ph} - q^2_1}
  \int \frac{d^2q_2}{(2\pi)^2 i}\,\frac{1}{\alpha_{\rm ph} - q^2_2}
  \ldots \nonumber\\
  \hspace{-0.3in}&&\times 
  \int \frac{d^2q_{2n + 1}}{(2\pi)^2 i}\,\frac{1}{\alpha_{\rm ph} - q^2_{2n + 1}}
  \frac{1}{\alpha_{\rm ph} - (q_1 + q_2 + \ldots + q_{2n + 1})^2}\Big\}.
\end{eqnarray}
In the momentum representation this proves relation (\ref{label3.22})
to second order in $\alpha_r(M^2)$ and to all orders in $\beta^2$:
\begin{eqnarray}\label{label3.26}
-i\,\tilde{\Delta}^{(2)}(p,\alpha_r(M^2))  = -i\,\tilde{\Delta}^{(2
)}(p,\alpha_{\rm ph}).
\end{eqnarray}
The proof of relation (\ref{label3.22}) to arbitrary orders in
$\alpha_r(M^2)$ and $\beta^2$ demands only patience and perseverance.

\subsection{Non--trivial finite corrections to the dimensional
  coupling constant $\alpha_{\rm ph}$}

As has been shown above quantum fluctuations around the trivial
vacuum, calculated to first order in $\alpha_r(M^2)$ and to all orders
in $\beta^2$, lead to the renormalization of the dimensional coupling
constant $\alpha_0(\Lambda^2)$, which reduces to the replacement
$\alpha_0(\Lambda^2) \to \alpha_{\rm ph}$, where $\alpha_{\rm ph}$ is
the physical (observable) dimensional coupling constant independent on
both the cut--off $\Lambda$ and the normalization scale $M$. In turn,
quantum fluctuations around the trivial vacuum, calculated to second
order in $\alpha_r(M^2)$ and to all orders in $\beta^2$, induce
non--trivial perturbative finite corrections to the physical coupling
constant $\alpha_{\rm ph}$.

The simplest correction of this kind is of order $O(\alpha^2_{\rm
  ph}\beta^2)$. It leads to the perturbative finite correction to the
soliton mass, coinciding with that calculated by Dashen {\it et al.}
\cite{RD74,RD75}.

Keeping only the terms of order $O(\beta^2)$ in (\ref{label3.25}) we get 
\begin{eqnarray}\label{label3.27}
   -i\,\tilde{\Delta}^{(2,1)}(p,\alpha_r(M^2)) &=&
  i\,\alpha^2_{\rm ph} \,
  \Big[\frac{(-i)}{\alpha_{\rm ph} - p^2 -i\,0}\Big]^2\,\beta^2
  \int \frac{d^2q_1}{(2\pi)^2 i}\,\frac{1}{(\alpha_{\rm ph} - q^2_1 - i\,0)^2} 
  =\nonumber\\
  &=& i\,\alpha_{\rm ph}\,\frac{\beta^2}{4\pi} \,
  \Big[\frac{(-i)}{\alpha_{\rm ph} - p^2 -i\,0}\Big]^2,
\end{eqnarray}
where we have taken into account that the integral over $q_1$ is equal
to $1/(4\pi \alpha_{\rm ph})$.

The two--point Green function calculated to second order in
$\alpha_{\rm ph}$ and to first order in $\beta^2$ is equal to
\begin{eqnarray}\label{label3.28}
  -i\,\tilde{\Delta}(p) = \frac{(-i)}{\alpha_{\rm ph} - p^2 -i\,0} + 
i\,\alpha_{\rm ph}\,\frac{\beta^2}{4\pi} \,
  \Big[\frac{(-i)}{\alpha_{\rm ph} - p^2 -i\,0}\Big]^2 = 
\frac{(-i)}{\alpha_{\rm eff} - p^2 -i\,0},
\end{eqnarray}
where $\alpha_{\rm eff}$ is defined by
\begin{eqnarray}\label{label3.29}
\alpha_{\rm eff} = \alpha_{\rm ph}\,\Big(1 -\frac{\beta^2}{4\pi}\Big).
\end{eqnarray}
The effective Lagrangian of the sine--Gordon model defining the
two--point Green function (\ref{label3.28}) to leading order in
$\beta^2$ takes the form
\begin{eqnarray}\label{label3.30}
{\cal L}(x) = \frac{1}{2}\,\partial_{\mu}\vartheta(x)
\partial^{\mu}\vartheta(x) +
\frac{\alpha_{\rm eff}}{\beta^2}\,
(\cos\beta\vartheta(x) - 1).
\end{eqnarray}
The soliton mass, calculated for the effective Lagrangian
(\ref{label3.30}), is equal to \cite{RD74,RD75}(see also \cite{FI6})
\begin{eqnarray}\label{label3.31}
  M_s = \frac{8\sqrt{\alpha_{\rm eff}}}{\beta^2} = 
\frac{8\sqrt{\alpha_{\rm ph}}}{\beta^2} - 
\frac{\sqrt{\alpha_{\rm ph}}}{\pi}.
\end{eqnarray}
This result coincides with that obtained by Dashen {\it et al.}
\cite{RD74,RD75}. 

We would like to remind that the finite correction
$-\,\sqrt{\alpha_{\rm ph}}/\pi$ has been interpreted in the literature
as a singularity of the sine--Gordon model at $\beta^2 = 8\pi$ (see
also \cite{LD78}).  However, as has been conjectured by Zamolodchikov
and Zamolodchikov \cite{AZ79}, such a singularity of the sine--Gordon
model is superficial and depends on the regularization and
renormalization procedure. This conjecture has been corroborated in
\cite{FI6}.

In our present analysis of the sine--Gordon model the finite
correction $-\,\sqrt{\alpha_{\rm ph}}/\pi$ is a perturbative one.  It
is valid only for $\beta^2 \ll 8\pi$ and introduces no singularity to
the sine--Gordon model at $\beta^2 = 8\pi$.

\subsection{Non--trivial momentum dependent corrections to the
  two--point Green function}

Quantum fluctuations around the trivial vacuum, calculated to second
order in $\alpha_{\rm ph}$ and to second order in $\beta^2$
inclusively, lead to the non--trivial momentum dependence of the
two--point Green function of the sine--Gordon field.  From
(\ref{label3.25}) we find the correction to the two--point Green
function to order $O(\alpha^2_{\rm ph}\beta^4)$ inclusively. This
reads
\begin{eqnarray}\label{label3.32}
  &&-i\,\tilde{\Delta}^{(2,2)}(p,\alpha_{\rm ph}) = i\,\alpha^2_{\rm ph}\,
  \Big[\frac{(-i)}{\alpha_{\rm ph} - p^2}\Big]^2\Big\{\beta^2 
  \int \frac{d^2q_1}{(2\pi)^2 i}\,\frac{1}{(\alpha_{\rm ph} - q^2_1)^2} \nonumber\\
  &&+ \beta^4\int \frac{d^2q_1}{(2\pi)^2 i}\,\frac{1}{\alpha_{\rm ph} - q^2_1}
  \int \frac{d^2q_2}{(2\pi)^2 i}\,\frac{1}{\alpha_{\rm ph} - q^2_2}\,
  \frac{1}{\alpha_{\rm ph} - (p - q_1 - q_2 )^2}\Big\}.
\end{eqnarray}
The momentum integral of the contribution of order $O(\beta^2)$ is
equal to $1/(4\pi \alpha_{\rm ph})$. For the calculation of the
momentum integrals of the term of order $O(\beta^4)$ we apply the
Feynman parameterization technique. This gives
\begin{eqnarray}\label{label3.33}
  &&\int \frac{d^2q_1}{(2\pi)^2 i}\,\frac{1}{\alpha_{\rm ph} - q^2_1}
  \int \frac{d^2q_2}{(2\pi)^2 i}\,\frac{1}{\alpha_{\rm ph} - q^2_2}\,
  \frac{1}{\alpha_{\rm ph} - (p - q_1 - q_2 )^2} =\nonumber\\
  &&= \frac{1}{16\pi^2}\int^1_0\int^1_0\int^1_0
  \frac{d\eta_1d\eta_2d\eta_3\delta(1 -
    \eta_1 - \eta_2 - \eta_3)}{\alpha_{\rm ph}(\eta_1\eta_2 + \eta_2\eta_3 + 
    \eta_3\eta_1)+ (-p^2\,)\eta_1\eta_2\eta_3} =\nonumber\\
  &&= \frac{1}{16\pi^2}\int^1_0d\eta\int^1_0d\xi\,\frac{1}{\alpha_{\rm ph}\eta + 
    (\alpha_{\rm ph} - p^2\,\eta)\,(1 - \eta)\,\xi\,(1 - \xi)} =\nonumber\\
  && = -\,\frac{1}{8\pi^2}\int^1_0\frac{d\eta }{
    \sqrt{(\alpha_{\rm ph} - p^2\eta)(1 - \eta)(4\alpha_{\rm ph}\eta 
+ (\alpha_{\rm ph} - p^2\eta)(1 - \eta)})}\nonumber\\
  &&\times\,
  {\ell n}\Big(\frac{\sqrt{(\alpha_{\rm ph} - p^2\eta)(1 - \eta)} +
    \sqrt{4\alpha_{\rm ph}\eta + (\alpha_{\rm ph} - p^2\eta)(1 - \eta)}}{
      \sqrt{(\alpha_{\rm ph} - p^2\eta)(1 - \eta)} - 
\sqrt{4\alpha_{\rm ph}\eta + (\alpha_{\rm ph} - p^2\eta)(1 - \eta)}}\Big).
\end{eqnarray}
We propose to analyse the behaviour of this integral in the asymptotic
regime $p^2 \to \infty$, where it can be calculated analytically. In
this limit the main contribution to the integral over $\eta$ comes
from the domain $\eta \sim \alpha_{\rm ph}/p^2$.  Therefore, the
integrand can be transcribed into the form
\begin{eqnarray}\label{label3.34}
  \hspace{-0.3in}&&\int \frac{d^2q_1}{(2\pi)^2 i}\,
  \frac{1}{\alpha_{\rm ph} - q^2_1}
  \int \frac{d^2q_2}{(2\pi)^2 i}\,\frac{1}{\alpha_{\rm ph} - q^2_2}\,
  \frac{1}{\alpha_{\rm ph} - (p - q_1 - q_2 )^2} =\nonumber\\ 
  \hspace{-0.3in}&&=-\,\frac{1}{8\pi^2}\int^1_0\frac{d\eta }{
    (\alpha_{\rm ph} - p^2\eta)}\,
  {\ell n}\Big(\frac{p^2\eta - \alpha_{\rm ph}}{\alpha_{\rm ph}\eta}\Big) = 
  \frac{1}{16\pi^2}\,\frac{1}{p^2}\,{\ell n}^2
\Big(\frac{p^2}{\alpha_{\rm ph}}\Big) + \ldots\,.
  \end{eqnarray}
  The correction to the two--point Green function of order
  $O(\alpha^2_{\rm ph}\beta^4)$ inclusively, taken in the asymptotic
  regime $p^2\to \infty$, is equal to
\begin{eqnarray}\label{label3.35}
  -i\,\tilde{\Delta}^{(2,2)}(p,\alpha_{\rm ph}) = i\,\alpha_{\rm ph}\,
  \Big[\frac{(-i)}{\alpha_{\rm ph} - p^2}\Big]^2\,\frac{\beta^2}{4\pi}\,
\Big[1 + \frac{\beta^2 }{4\pi}\,\frac{\alpha_{\rm ph}}{p^2}\,
{\ell n}^2\Big(\frac{p^2}{\alpha_{\rm ph}}\Big) + \ldots\Big].
\end{eqnarray}
Hence, the two--point Green function of the sine--Gordon field,
accounting for the contributions of order $O(\alpha^2_{\rm
  ph}\beta^4)$ inclusively, takes the form
\begin{eqnarray}\label{label3.36}
  \tilde{\Delta}^{-1}(p) = \alpha_{\rm ph}
  \Big[1 -\frac{\beta^2}{4\pi}- \frac{\beta^4 }{16\pi^2}\,
  \frac{\alpha_{\rm ph}}{p^2}\,{\ell n}^2
  \Big(\frac{p^2}{\alpha_{\rm ph}}\Big) + \ldots \Big] - p^2.
\end{eqnarray}
This shows that (i) all divergences can be removed by the
renormalization of the dimensional coupling constant
$\alpha_0(\Lambda^2)$, (ii) the renormalized expressions are defined
in terms of the physical coupling constant $\alpha_{\rm ph}$, (iii)
higher order corrections in $\alpha_{\rm ph}$ introduce a non--trivial
momentum dependence and (iv) in the asymptotic limit $p^2 \to \infty$
the two--point Green function of the sine--Gordon field behaves as
$\tilde{\Delta}(p) \to 1/( - p^2)$. Such a behaviour is confirmed by
the analysis of the two--point Green functions with the
Callan--Symanzik equation (see Section 4).

\subsection{Physical renormalization of the sine--Gordon model}

Using the results obtained above we can formulate a procedure for the
renormalization of the sine--Gordon model dealing with physical
parameters only.  Starting with the Lagrangian (\ref{label1.1}) and
making a renormalization at the normalization scale $M^2 = \alpha_{\rm
  ph}$ we deal with physical parameters only
\begin{eqnarray}\label{label3.37}
\alpha_{\rm ph} = Z^{-1}_1(\beta^2, \alpha_{\rm ph};\Lambda^2)\,
\alpha_0(\Lambda^2),
\end{eqnarray}
where the renormalization constant $Z_1( \beta^2, \alpha_{\rm ph};
\Lambda^2)$ is equal to
\begin{eqnarray}\label{label3.38}
Z_1(\beta^2,\alpha_{\rm ph}; \Lambda^2) =
\Big(\frac{\Lambda^2}{\alpha_{\rm ph}}\Big)^{\beta^2/8\pi}.
\end{eqnarray}
The renormalized Lagrangian is defined by
\begin{eqnarray}\label{label3.39}
{\cal L}(x) = \frac{1}{2}\,(\partial_{\mu}\vartheta(x)
\partial^{\mu}\vartheta(x) + \frac{\alpha_{\rm ph}}{\beta^2}\,
(\cos\beta\vartheta(x) - 1) + (Z_1 - 1)\,\frac{\alpha_{\rm ph}}{\beta^2}\,
(\cos\beta\vartheta(x) - 1)
\end{eqnarray}
with the renormalization constant given by Eq.(\ref{label3.37}). From
the relation (\ref{label3.20}) at $M^2 = \alpha_{\rm ph}$ one can
obtain that
\begin{eqnarray}\label{label3.40}
 \alpha_r(\alpha_{\rm ph}) = \alpha_{\rm ph}.
\end{eqnarray}
The calculation of perturbative corrections to the two--point Green
function of the sine--Gordon model shows that the first order
correction in $\alpha_{\rm ph}$ vanishes in accordance with
Eq.(\ref{label3.17}). Non--trivial corrections appear only to second
and higher orders in $\alpha_{\rm ph}$.

\section{Renormalization group analysis}
\setcounter{equation}{0}

In this Section we discuss the renormalization group approach
\cite{IZ80}--\cite{SW96} to the renormalization of the sine--Gordon
model. We apply the Callan--Symanzik equation to the analysis of the
Fourier transform of the two--point Green function of the sine--Gordon
field.

The Callan--Symanzik equation for the Fourier transform of the
two--point Green function of the sine--Gordon field
(\ref{label3.5}), which we denote below as $-i\,\tilde{\Delta}(p;
\alpha_r(M^2), \beta^2)$, is equal to \cite{IZ80}
\begin{eqnarray}\label{label4.1}
  \Big[- p\cdot \frac{\partial}{\partial p} + \beta(\alpha_r(M^2),\beta^2)\,
  \frac{\partial}{\partial \alpha_r(M^2)} - 2\Big]\tilde{\Delta}(p;
  \alpha_r(M^2), \beta^2) = F(0,p;\alpha_r(M^2),\beta^2),
\end{eqnarray}
where $\beta(\alpha_r(M^2),\beta^2)$ is the Gell--Mann--Low
function
\begin{eqnarray}\label{label4.2}
 M\,\frac{\partial \alpha_r(M^2)}{\partial M} =  \beta(\alpha_r(M^2),\beta^2). 
\end{eqnarray}
The term $\gamma(\alpha_r(M^2),\beta^2)$ \cite{IZ80}, describing an
anomalous dimension of the sine--Gordon field, does not appear in the
Callan--Symanzik equation (\ref{label4.1}) due to unrenormalizability
of the sine--Gordon field $\vartheta(x)$. The r.h.s. of
(\ref{label4.1}) is defined by
\begin{eqnarray}\label{label4.3}
  F(0,p;\alpha_r(M^2),\beta^2) = \int\!\!\!\int  
  d^2x d^2y\,e^{\textstyle+ip \cdot x}\,
  \langle 0|
  {\rm T}\Big(\Theta^{\mu}_{\mu}(y)\vartheta(x)\vartheta(0)\,
e^{\textstyle i\int d^2y\,{\cal L}_{\rm int}(y)}\Big)|0\rangle_c,
\end{eqnarray}
where ${\cal L}_{\rm int}(y)$ is equal to \cite{FI6}
\begin{eqnarray}\label{labe4.4}
  {\cal L}_{\rm int}(y) = \frac{\alpha_0}{\beta^2}\,(\cos\beta\vartheta(x) - 1).
\end{eqnarray}
Then, $\Theta^{\mu}_{\mu}(y)$ is the trace of the energy--momentum
tensor $\Theta_{\mu\nu}(x)$. For a (pseudo)scalar field
$\vartheta(x)$, described by the Lagrangian ${\cal L}(x)$, it is
defined by \cite{IZ80}
\begin{eqnarray}\label{label4.5}
  \Theta_{\mu\nu}(x) = \frac{\partial {\cal L}(x)}{\partial^{\mu}\vartheta(x)}\,
\partial_{\nu}\vartheta(x) - g_{\mu\nu}\,{\cal L}(x).
\end{eqnarray}
Using the Lagrange equation of motion one can show that 
\begin{eqnarray}\label{label4.6}
  \partial^{\mu}\Theta_{\mu\nu}(x) = 0.
\end{eqnarray}
For the sine--Gordon model the energy--momentum tensor
$\Theta_{\mu\nu}(x)$ reads

\begin{eqnarray}\label{label4.7}
  \Theta_{\mu\nu}(x) = \partial_{\mu}\vartheta(x) \partial_{\nu}\vartheta(x) - 
g_{\mu\nu}\,\Big[\frac{1}{2}\partial_{\lambda}\vartheta(x) \partial^{\lambda}
\vartheta(x) + \frac{\alpha_0}{\beta^2}\,(\cos\beta \vartheta(x) - 1)\Big].
\end{eqnarray}
The trace of the energy--momentum tensor $\Theta_{\mu\nu}(x)$ is equal to
\begin{eqnarray}\label{label4.8}
  \Theta^{\mu}_{\mu}(x) = -\,\frac{2\alpha_0}{\beta^2}\,
(\cos\beta \vartheta(x) - 1) = 2\,V[\vartheta(x)],
\end{eqnarray}
where $V[\vartheta(x)]$ is the potential density functional of the
sine--Gordon field $\vartheta(x)$.

Since the trace of the energy--momentum tensor is proportional to the
potential energy density, the Fourier transform
$F(0,p;\alpha_r(M^2),\beta^2)$ can be related to the two--point Green
function as
\begin{eqnarray}\label{label4.9}
  && F(0,p;\alpha_r(M^2),\beta^2) = 2\,\alpha_r(M^2)\,
\frac{\partial}{\partial \alpha_r(M^2)} 
  \tilde{\Delta}(p;\alpha_r(M^2),\beta^2),
\end{eqnarray}
where we have used the definition of the trace $\Theta^{\mu}_{\mu}(y)$
of the energy--momentum tensor Eq.(\ref{label4.8}) and the relation
$\alpha_0 = \alpha_r(M^2)\,Z_1(\beta^2, M^2;
\Lambda^2)$.

Substituting (\ref{label4.9}) into (\ref{label4.1}) we arrive
at the Callan--Symanzik equation for the Fourier transform of the
two--point Green function of the sine--Gordon field
\begin{eqnarray}\label{label4.10}
  \Big[- p^2\frac{\partial}{\partial p^2} + \Big(\frac{1}{2}\,
\beta(\alpha_r(M^2),\beta^2) - 
  \alpha_r(M^2)\Big)\,\frac{\partial}{\partial \alpha_r(M^2)} - 
  1\Big]\tilde{\Delta}(p^2;
  \alpha_r(M^2), \beta^2) = 0,
\end{eqnarray}
where we have taken into account that $\tilde{\Delta}(p;
\alpha_r(M^2), \beta^2)$ should depend on $p^2$ due to Lorentz
covariance.

For the solution of (\ref{label4.10}) we have to determine the
Gell--Mann--Low function (\ref{label4.2}). For the coupling constant
$\alpha_r(M^2)$, defined by (\ref{label3.20}), the Gell--Mann--Low
function is
\begin{eqnarray}\label{label4.11}
 \beta(\alpha_r(M^2),\beta^2) = \frac{\tilde{\beta}^2}{4\pi}\,\alpha_r(M^2),
\end{eqnarray}
where $\tilde{\beta}^2 = \beta^2/(1 + \beta^2/8\pi)$
(\ref{label3.20}). This gives the Callan--Symanzik equation 
\begin{eqnarray}\label{label4.12}
  \Big[p^2\frac{\partial}{\partial p^2} +  \Big(1 - 
  \frac{\tilde{\beta}^2}{8\pi}\Big)\,\alpha_r(M^2)\,
\frac{\partial}{\partial \alpha_r(M^2)} + 
  1\Big]\tilde{\Delta}(p^2;
  \alpha_r(M^2), \beta^2) = 0.
\end{eqnarray}
Setting $\tilde{\Delta}(p^2;
\alpha_r(M^2), \beta^2)= D(p^2; \alpha_r(M^2), \beta^2)/p^2$ we get
\begin{eqnarray}\label{label4.13}
  \Big[ p^2\frac{\partial}{\partial p^2} + 
  \Big(1 - \frac{\tilde{\beta}^2}{8\pi}\Big)\,\alpha_r(M^2)\,
  \frac{\partial}{\partial \alpha_r(M^2)}\Big]D(p^2;
  \alpha_r(M^2), \beta^2) = 0.
\end{eqnarray}
Due to dimensional consideration the function $D(p^2; \alpha_r(M^2),
\beta^2)$ should be dimensionless, depending on the dimensionless
variables $\tilde{p}^2 = p^2/M^2$ and $\tilde{\alpha} =
\alpha_r(M^2)/M^2$, where $M$ is a normalization scale. This gives
\begin{eqnarray}\label{label4.14}
  \Big[ \tilde{p}^2\frac{\partial}{\partial \tilde{p}^2} + 
  \Big(1 - \frac{\tilde{\beta}^2}{8\pi}\Big)\,\tilde{\alpha}\,
  \frac{\partial}{\partial \tilde{\alpha}}\Big]D(\tilde{p}^2;
  \tilde{\alpha}, \beta^2) = 0.
\end{eqnarray}
According to the general theory of partial differential equations of
first order \cite{RC62}, the solution of (\ref{label4.14}) is an
arbitrary function of the integration constant
\begin{eqnarray}\label{label4.15}
  C = \frac{\tilde{\alpha}}{\tilde{p}^2}\,(\tilde{p}^2)^{\tilde{\beta}^2/8\pi},
\end{eqnarray}
which is the solution of the characteristic differential equation
\begin{eqnarray}\label{label4.16}
 \Big(1- \frac{\tilde{\beta}^2}{8\pi}\Big)\frac{d \tilde{p}^2}{\tilde{p}^2} = \frac{d \tilde{\alpha}}{\tilde{\alpha}}. 
\end{eqnarray}
Hence, the Fourier transform of the two--point Green
function of the sine--Gordon field is equal to
\begin{eqnarray}\label{label4.17}
  \tilde{\Delta}(p^2;
  \alpha_r(M^2), \beta^2) = \frac{1}{p^2}\,D\Big[\frac{\alpha_r(M^2)}{p^2}\,
\Big(\frac{p^2}{M^2}\Big)^{\tilde{\beta}^2/8\pi}\Big].
\end{eqnarray}
The argument of the $D$--function can be
expressed in terms of the running coupling constant $\alpha_r(p^2)$:
\begin{eqnarray}\label{label4.18}
  \alpha_r(p^2) = \alpha_r(M^2)\Big(\frac{p^2}{M^2}\Big)^{\tilde{\beta}^2/8\pi} = 
\alpha_{\rm ph}\Big(\frac{M^2}{\alpha_{\rm ph}}\Big)^{\tilde{\beta}^2/8\pi}
 \Big(\frac{p^2}{M^2}\Big)^{\tilde{\beta}^2/8\pi} = \alpha_{\rm ph}
\Big(\frac{p^2}{\alpha_{\rm ph}}\Big)^{\tilde{\beta}^2/8\pi}. 
\end{eqnarray}
The solution of the Callan--Symanzik equation for the Fourier
transform of the two--point Green function of the sine--Gordon field
is
\begin{eqnarray}\label{label4.19}
  \tilde{\Delta}(p^2; \alpha_{\rm ph}, \beta^2) = \frac{1}{p^2}\,
D\Big[\frac{\alpha_r(p^2)}{p^2}\,\Big].
\end{eqnarray}
This proves that the total renormalized two--point Green function of
the sine--Gordon field depends on the physical coupling constant
$\alpha_{\rm ph}$ only. 

A perturbative calculation of the two--point Green function, carried
out in Section 3, gives the following expression for the function
$D[\alpha_r(p^2)/p^2]$ in the asymptotic region $p^2 \to \infty$:
\begin{eqnarray}\label{label4.20}
   D\Big[\frac{\alpha_r(p^2)}{p^2}\,\Big] = 
\Big\{\frac{\alpha_{\rm ph}}{p^2}\Big[1 -\frac{\beta^2}{4\pi} - 
\frac{\beta^4 }{16\pi^2}\,\frac{\alpha_{\rm ph}}{p^2}\,{\ell n}^2
\Big(\frac{p^2}{\alpha_{\rm ph}}\Big) + \ldots\Big] - 1\Big\}^{-1}.
\end{eqnarray}
Unfortunately, this is not able to reproduce the non--perturbative
expression of the two--point Green function $ \tilde{\Delta}(p^2;
\alpha_{\rm ph}, \beta^2)$.

\section{Renormalization of Gaussian fluctuations around solitons}
\setcounter{equation}{0}

We apply the renormalization procedure expounded above to the
calculation of the contribution of quantum fluctuations around a
soliton solution. We start with the partition function
\begin{eqnarray}\label{label5.1}
Z_{\rm SG} &=& \int {\cal D}\vartheta\,\exp\Big\{i\int
d^2x\,\Big[\frac{1}{2}\,\partial_{\mu}\vartheta(x)
\partial^{\mu}\vartheta(x) +
\frac{\alpha_0}{\beta^2}\,(\cos\beta\vartheta(x) -
1)\Big]\Big\}=\nonumber\\ &=&\int {\cal D}\vartheta\,\exp\Big\{i\int
d^2x\,{\cal L}[\vartheta(x)]\Big\}.
\end{eqnarray}
Following Dashen {\it et al.}  \cite{RD74,RD75} (see also \cite{JR70})
we treat the quantum fluctuations of the sine--Gordon field
$\vartheta(x)$ around the classical solution $\vartheta(x) =
\vartheta_{\rm c\ell}(x) + \varphi(x)$, where $\varphi(x)$ is the
field fluctuating around $\vartheta_{\rm c\ell}(x)$, the single
soliton solution of the classical equation of motion
\begin{eqnarray}\label{label5.2}
\Box\vartheta_{\rm c\ell}(x) +
\frac{\alpha_0}{\beta}\,\sin\beta\vartheta_{\rm c\ell}(x) = 0
\end{eqnarray}
equal to \cite{RD74,LD78,JR70}
\begin{eqnarray}\label{label5.3}
\vartheta_{\rm c\ell}(x) = \frac{4}{\beta}\,\arctan(\exp(
\sqrt{\alpha_0}\,\gamma (x^1 - ux^0)) =
\frac{4}{\beta}\,\arctan(\exp(\sqrt{\alpha_0}\,\sigma)), 
\end{eqnarray}
where $u$ is the velocity of the soliton, $\sigma = \gamma (x^1 -
ux^0)$ and $\gamma = 1/\sqrt{1 - u^2}$\,\footnote{In analogy with the
  ``spatial'' variable $\sigma$ we can define the ``time'' variable
  for the soliton moving with velocity $u$ as $\tau = \gamma (x^0 - u
  x^1)$. In variables $(\tau,\sigma)$ an infinitesimal element of the
  2--dimensional volume $d^2x$ is equal to $d^2x = d\tau d\sigma$ and
  the d'Alembert operator $\Box$ is  defined by $\Box =
  \partial^2/\partial \tau^2 - \partial^2/\partial \sigma^2$.}.

Substituting $\vartheta(x) = \vartheta_{\rm c\ell}(x) + \varphi(x)$
into the exponent of the integrand of (\ref{label5.1}) and using the
equation of motion for the soliton solution $\vartheta_{\rm c\ell}(x)$
we get
\begin{eqnarray}\label{label5.4}
&&Z_{\rm SG} = \exp\Big\{i\int d^2x\,{\cal L}[\vartheta_{\rm
c\ell}(x)]\Big\}\int {\cal D}\varphi\,\exp\Big\{i\int
d^2x\,\Big[\frac{1}{2}\,\partial_{\mu}\varphi(x)
\partial^{\mu}\varphi(x)\nonumber\\ &&+
\frac{\alpha_0}{\beta^2}\,\sin\beta\vartheta_{\rm c\ell}(x)\,(\beta
\varphi(x) - \sin \beta \varphi(x)) + \frac{\alpha_0}{\beta^2}\,
\cos\beta\vartheta_{\rm c\ell}(x)(\cos\beta\varphi(x) - 1)\Big]\Big\}.
\end{eqnarray}
Substituting (\ref{label5.3}) into (\ref{label5.4}) we obtain
\begin{eqnarray}\label{label5.5}
\hspace{-0.3in}{\cal L}[\varphi(x)] &=&
\frac{1}{2}\,\partial_{\mu}\varphi(x) \partial^{\mu}\varphi(x) +
\frac{\alpha_0}{\beta^2}\,(\cos\beta\varphi(x) - 1)\nonumber\\
\hspace{-0.3in}&-&
\frac{2\alpha_0}{\beta^2}\,\frac{1}{\cosh^2(\sqrt{\alpha_0}\,\gamma
(x^1 - u x^0))}(\cos\beta\varphi(x) - 1)\nonumber\\ \hspace{-0.3in}&-&
 \frac{2\alpha_0}{\beta^2}\,\frac{\sinh(\sqrt{\alpha_0}\,\gamma(x^1 -
u x^0))}{\cosh^2(\sqrt{\alpha_0}\,\gamma (x^1 - u x^0))}\,(\beta
\varphi(x) - \sin\beta\varphi(x)).
\end{eqnarray}
In terms of $\sigma$ and $\tau$ the exponent of the partition function
(\ref{label5.4}) reads
\begin{eqnarray}\label{label5.6}
\hspace{-0.3in}&&Z_{\rm SG} = \exp\Big\{i\int d^2x\,{\cal
L}[\vartheta_{\rm c\ell}(x)]\Big\}\nonumber\\ \hspace{-0.3in}&& \int
{\cal D}\varphi\,\exp\Big\{i\int d\tau
d\sigma\,\Big[\frac{1}{2}\,\partial_{\mu}\varphi(\tau,\sigma)
\partial^{\mu}\varphi(\tau,\sigma) + \frac{\alpha_0}{\beta^2}\,(\cos
\beta \varphi(\tau,\sigma) - 1)\nonumber\\ \hspace{-0.3in}&& -
\frac{2\alpha_0}{\beta^2}\,\frac{1}{\cosh^2(\sqrt{\alpha_0}\sigma)}
(\cos\beta\varphi(x) - 1) - \frac{2\alpha_0}{\beta^2}\,
\frac{\sinh(\sqrt{\alpha_0}\sigma)}{\cosh^2(\sqrt{\alpha_0}\sigma)}\,
(\beta \varphi(\tau, \sigma) - \sin\beta\varphi(\tau,
\sigma))\Big]\Big\},\quad\quad
\end{eqnarray}
where we have denoted
\begin{eqnarray}\label{label5.7}
\partial_{\mu}\varphi(\tau,\sigma) \partial^{\mu}\varphi(\tau,\sigma)
= \Big(\frac{\partial \varphi(\tau,\sigma)}{\partial \tau}\Big)^2 -
\Big(\frac{\partial \varphi(\tau,\sigma)}{\partial \sigma}\Big)^2.
\end{eqnarray}
The equation of motion for the fluctuating field
$\varphi(\tau,\sigma)$ is equal to
\begin{eqnarray}\label{label5.8}
  \Box \varphi(\tau,\sigma) + \frac{\alpha_0}{\beta}\,\sin\beta
  \varphi(\tau,\sigma) &=& +
  2\,\frac{\alpha_0}{\beta}\,\frac{1}{\cosh^2(\sqrt{\alpha_0}\sigma)}\,
  \sin\beta\varphi(\tau,\sigma)\nonumber\\ && -
  2\,\frac{\alpha_0}{\beta}\,
  \frac{\sinh(\sqrt{\alpha_0}\sigma)}{\cosh^2(\sqrt{\alpha_0}\sigma)}\,(1
  - \cos\beta\varphi(\tau,\sigma)).
\end{eqnarray}
Dealing with Gaussian fluctuations only \cite{RD74,RD75} and keeping
the squared terms in the Lagrangian (\ref{label5.5}) in the
fluctuating field $\varphi(x)$ expansion only, we transcribe the
partition function (\ref{label5.6}) into the form
\begin{eqnarray}\label{label5.9}
\hspace{-0.3in}&&Z_{\rm SG} = \exp\Big\{i\int d^2x\,{\cal
L}[\vartheta_{\rm c\ell}(x)]\Big\}\nonumber\\ \hspace{-0.3in}&& \times
\int {\cal D}\varphi\,\exp\Big\{\,-\,i\,\frac{1}{2}\int d\tau
d\sigma\,\varphi(\tau,\sigma) \Big[\Box + \alpha_0 -
\frac{2\alpha_0}{\cosh^2(\sqrt{\alpha_0}\sigma)}\Big]
\varphi(\tau,\sigma) \Big\}.
\end{eqnarray}
It is seen that $\sqrt{\alpha_0}$ has the distinct meaning of the mass
of the quanta of the Klein--Gordon field $\varphi(\tau,\sigma)$
coupled to an external force described by a scalar
potential\,\footnote{The parameter $\alpha_0$ should enter with the
  imaginary correction $\alpha_0 \to \alpha_0 -i\,0$. This is required
  by the convergence of the path integral \cite{MP95a}.}.

Integrating over the fluctuating field $\varphi(\tau,\sigma)$ we
transcribe the r.h.s. of (\ref{label5.9}) into the form
\begin{eqnarray}\label{label5.10}
\hspace{-0.3in}&&Z_{\rm SG} = \exp\Big\{i\int d^2x\,{\cal
L}[\vartheta_{\rm c\ell}(x)] + i\delta {\cal S}[\vartheta_{\rm
c\ell}]\Big\}.
\end{eqnarray}
We have denoted
\begin{eqnarray}\label{label5.11}
  \hspace{-0.3in} \exp\{i\delta {\cal S}[\vartheta_{\rm
c\ell}]\} &=& \exp\{ i\int d^2x\,\delta
  {\cal L}_{\rm eff}[\vartheta_{\rm c\ell}(x)]\} = \sqrt{\frac{{\rm Det}( \Box + \alpha_0)}{\displaystyle 
      {\rm Det}\Big( \Box + \alpha_0 -
      \frac{2\alpha_0}{\cosh^2(\sqrt{\alpha_0}\sigma)}\Big)}} = \nonumber\\
  \hspace{-0.3in}&=&
  \exp\Big\{-\frac{1}{2}\sum_n{\ell n}\,\lambda_n + \frac{1}{2}
  \int d^2x\int 
  \frac{d^2p}{(2\pi)^2}\,{\ell n}(\alpha_0 - p^2)\Big\},
\end{eqnarray}
where $p$ is a 1+1--dimensional momentum. The second term in the
exponent corresponds to the subtraction of the vacuum contribution.
The effective action, caused by fluctuations around a soliton
solution, is defined by
\begin{eqnarray}\label{label5.12}
  \delta {\cal S}[\vartheta] = \int d^2x\,
\delta {\cal L}_{\rm eff}[\vartheta_{\rm c\ell}(x)] =
  i\,\frac{1}{2}\sum_n{\ell n}\,\lambda_n + \frac{1}{2}\int d^2x\int 
  \frac{d^2p}{(2\pi)^2i}\,{\ell n}(\alpha_0 - p^2),
\end{eqnarray}
where $\lambda_n$ are the eigenvalues of the equation
\begin{eqnarray}\label{label5.13}
  \Big(\Box + \alpha_0 -
  \frac{2\alpha_0}{\cosh^2(\sqrt{\alpha_0}\sigma)}\Big)\,\varphi_n(\tau,\sigma) =
  \lambda_n\,\varphi_n(\tau,\sigma)
\end{eqnarray}
and $\varphi_n(\tau,\sigma)$ are eigenfunctions. The quantum number
$n$ can be both discrete and continuous. This implies that the product
over $n$ in (\ref{label5.11}) should contain both the summation over
the discrete values of the quantum number $n$ and integration over the
continuous ones.
 
According to the Fourier method \cite{NN65}, the solution of equation
(\ref{label5.13}) should be taken in the form
\begin{eqnarray}\label{label5.14}
\varphi_n(\tau,\sigma)= e^{\textstyle\,- i\,\omega\tau
}\,\psi_n(\sigma),
\end{eqnarray}
where $-\,\infty \le \omega \le +\,\infty$ and $\psi_n(\sigma)$ is a
complex function\,\footnote{Since $\varphi_n(\tau,\sigma)$ is a real
  field, we have to take the real part of the solution
  (\ref{label5.14}) only, i.e. $\varphi(\tau,\sigma) = {\cal
    R}e\,(e^{\textstyle\,- i\,\omega\tau}\, \psi(\sigma))$. Though
  without loss of generality one can also use complex eigenfunctions
  \cite{RD75,LD78,JR70}.}.

Substituting (\ref{label5.14}) into (\ref{label5.13}) we get
\begin{eqnarray}\label{label5.15}
\Big(\frac{d^2}{d \sigma^2} + k^2 +
\frac{2\alpha_0}{\cosh^2(\sqrt{\alpha_0}\,\sigma)}\Big)\,
\psi_n(\tau,\sigma) = 0,
\end{eqnarray}
where we have denoted 
\begin{eqnarray}\label{label5.16}
k^2 = \lambda_n + \omega^2 - \alpha_0.
\end{eqnarray}
This defines eigenvalues $\lambda_n$ as functions of $\omega$ and $k$
\begin{eqnarray}\label{label5.17}
\lambda_n= \alpha_0 - \omega^2 + k^2.
\end{eqnarray}
The parameter $k$ has the meaning of a spatial momentum $-\infty < k <
+ \infty$. The solutions of equation (\ref{label5.15})
are\,\footnote{The solutions of the equation (\ref{label5.15}) are
  well--known \cite{JR70} (see also \cite{RD75,LD78}). Nevertheless,
  we adduce the solution of this equation in the Appendix.}
\begin{eqnarray}\label{label5.18}
\psi_b(\sigma) &=& \sqrt{\frac{\sqrt{\alpha_0}}{2}}\,
\frac{1}{\cosh(\sqrt{\alpha_0}\sigma)},\nonumber\\ \psi_k(\sigma) &=&
\frac{i}{\sqrt{2\pi}}\,\frac{-ik +
\sqrt{\alpha_0}\,\tanh(\sqrt{\alpha_0}\sigma)}{\sqrt{k^2 +
\alpha_0}}\,e^{\textstyle\,+ ik\sigma},
\end{eqnarray}
where the eigenfunction $\psi_b(\sigma)$ has eigenvalues $\lambda_n =
-\omega^2$ and the eigenfunctions $\psi_k(\sigma)$ have eigenvalues
$\lambda_n = \alpha_0 - \omega^2 + k^2$.  In the asymptotic region
$\sigma \to \infty$ the function $\psi_k(\sigma)$ behaves as
\begin{eqnarray}\label{label5.19}
  \psi_k(\sigma) \to \frac{1}{\sqrt{2\pi}}\,e^{\textstyle + ik\sigma + 
i\,\frac{1}{2}\,\delta(k)},
\end{eqnarray}
where $\delta(k)$ is a phase shift defined by \cite{JR70}
\begin{eqnarray}\label{label5.20}
\delta(k) = 2\arctan\frac{\sqrt{\alpha_0}}{k}.
\end{eqnarray}
The solutions (\ref{label5.18}) satisfy the completeness condition
\cite{JR70} (see Appendix)
\begin{eqnarray}\label{label5.21}
\int^{+\infty}_{-\infty}dk\,\psi^*_k(\sigma\,'\,)\psi_k(\sigma) +
\psi_b(\sigma\,'\,)\psi_b(\sigma) = \delta(\sigma\,' - \sigma).
\end{eqnarray}
The fluctuating field $\varphi(\tau,\sigma)$ is equal to (see ({\rm
A}.12))
\begin{eqnarray}\label{label5.22}
  \hspace{-0.3in}\varphi_{\omega b}(\tau,\sigma) &=& \frac{1}{\sqrt{2\pi}}\,e^{\textstyle -i\omega \tau}\psi_b(\sigma)  = \frac{1}{\sqrt{2\pi}}\,
\sqrt{\frac{\sqrt{\alpha_0}}{2}}\,
  \frac{1}{\cosh(\sqrt{\alpha_0}\sigma)}\,e^{\textstyle -i\omega \tau},\nonumber\\
   \hspace{-0.3in}\varphi_{\omega k}(\tau,\sigma) &=& \frac{1}{\sqrt{2\pi}}\,e^{\textstyle -i\omega \tau}\psi_k(\sigma) = \frac{i}{2\pi}\,\frac{-ik +
    \sqrt{\alpha_0}\,\tanh(\sqrt{\alpha_0}\sigma)}{\sqrt{k^2 +
      \alpha_0}}\,e^{\textstyle\,-i\omega \tau + ik\sigma}. 
\end{eqnarray}
In terms of the eigenvalues $\lambda_n = -\,\omega^2$ and $\lambda_n =
\alpha_0 - \omega^2 + k^2$ and eigenfunctions (\ref{label5.22}) the
effective action $\delta {\cal S}[\vartheta_{\rm c\ell}]$ is
determined by
\begin{eqnarray}\label{label5.23}
\hspace{-0.3in}&&\delta {\cal S}[\vartheta_{\rm c\ell}] = -
\frac{1}{2}\int d^2x\,\int^{+\infty}_{-\infty}\frac{d\omega}{2\pi
i}\int^{+\infty}_{-\infty}dk\,|\psi_k(x)|^2\,{\ell n}(\alpha_0 -
\omega^2 + k^2)\nonumber\\
\hspace{-0.3in}&& - \frac{1}{2}\int
d^2x\,\int^{+\infty}_{-\infty}\frac{d\omega}{2\pi
i}\,|\psi_b(x)|^2\,{\ell n}(- \omega^2)  + \frac{1}{2}\int d^2x\int 
  \frac{d^2p}{(2\pi)^2i}\,{\ell n}(\alpha_0 - p^2).
\end{eqnarray}
Using the explicit expressions for the eigenfunctions $\psi_k(x)$ and
$\psi_b(x)$ we reduce the r.h.s. of (\ref{label5.23}) to the form
\begin{eqnarray}\label{label5.24}
\hspace{-0.7in}&&\delta {\cal S}[\vartheta_{\rm c\ell}] = -
\frac{1}{2}\int d^2x\,\int^{+\infty}_{-\infty}\frac{d\omega}{2\pi i
}\int^{+\infty}_{-\infty}\frac{dk}{2\pi}\,\frac{k^2 +
\alpha_0\tanh^2(\sqrt{\alpha_0}\sigma)}{k^2 + \alpha_0}\,{\ell
n}(\alpha_0 - \omega^2 + k^2)\nonumber\\
\hspace{-0.7in}&&- \frac{1}{2}\int
d^2x\,\int^{+\infty}_{-\infty}\frac{d\omega}{2\pi
i}\,\frac{\sqrt{\alpha_0}}{2}\,\frac{1}{\cosh^2(\sqrt{\alpha_0}\sigma)}
\,{\ell n}(- \omega^2) + \frac{1}{2}\int d^2x\int 
  \frac{d^2p}{(2\pi)^2i}\,{\ell n}(\alpha_0 - p^2).
\end{eqnarray}
Introducing the notation
\begin{eqnarray}\label{label5.25}
\frac{1}{\cosh^2(\sqrt{\alpha_0}\sigma)} = \frac{1}{2}\,(1 - \cos\beta
\vartheta_{\rm c\ell}(x)) = 
\frac{\beta^2}{2\alpha_0}\,V[\vartheta_{\rm c\ell}(x)]
\end{eqnarray}
we obtain the effective Lagrangian $\delta {\cal L}_{\rm
  eff}[\vartheta_{\rm c\ell}(x)]$. It is equal to\,\footnote{In the
  contribution of the mode $\lambda_n = - \omega^2$ we have used the
  integral representation \[\frac{1}{2\sqrt{\alpha_0}} =
  \int^{+\infty}_{-\infty}\frac{dk}{2\pi}\,\frac{1}{k^2 +
    \alpha_0}.\]}
\begin{eqnarray}\label{label5.26}
 \delta {\cal L}_{\rm eff}[\vartheta_{\rm c\ell}(x)] &=&
  -\frac{1}{4}\beta^2V[\vartheta_{\rm
    c\ell}(x)]\int^{+\infty}_{-\infty}\frac{d\omega}{2\pi i
  }\int^{+\infty}_{-\infty}\frac{dk}{2\pi}\frac{1}{k^2 +
    \alpha_0} [{\ell n}(- \omega^2) - {\ell n}(\alpha_0 - \omega^2 +
  k^2)]\nonumber\\
 &-&
  \frac{1}{2} \int^{+\infty}_{-\infty}\frac{d\omega}{2\pi i
  }\int^{+\infty}_{-\infty}\frac{dk}{2\pi} {\ell n}(\alpha_0 - \omega^2 + k^2) 
+ \frac{1}{2}\int 
  \frac{d^2p}{(2\pi)^2i} {\ell n}(\alpha_0 - p^2). 
\end{eqnarray}
The two last terms cancel each other. This gives
\begin{eqnarray}\label{label5.27}
\hspace{-0.3in} &&\delta {\cal L}_{\rm eff}[\vartheta_{\rm c\ell}(x)] =
  -\frac{1}{4}\beta^2V[\vartheta_{\rm
    c\ell}(x)]\int^{+\infty}_{-\infty}\frac{d\omega}{2\pi i
  }\int^{+\infty}_{-\infty}\frac{dk}{2\pi}\frac{1}{k^2 +
    \alpha_0} [{\ell n}(- \omega^2) - {\ell n}(\alpha_0 - \omega^2 +
  k^2)].\nonumber\\
\hspace{-0.3in} &&
\end{eqnarray}
After the integration by parts over $\omega$ the effective Lagrangian
$\delta {\cal L}_{\rm eff}[\vartheta_{\rm c\ell}(x)]$ is defined by
\begin{eqnarray}\label{label5.28}
\delta {\cal L}_{\rm eff}[\vartheta_{\rm c\ell}(x)] = 
\frac{1}{2}\,\beta^2V[\vartheta_{\rm
    c\ell}(x)]\int^{+\infty}_{-\infty}\frac{dk}{2\pi}\int^{+\infty}_{-\infty}
\frac{d\omega}{2\pi
    i }\frac{1}{\alpha_0 - \omega^2 + k^2 - i\,0}.
\end{eqnarray}
The appearance of the imaginary correction $-i\,0$ is caused by the
convergence of the path integral (\ref{label5.9}) \cite{MP95a}.

Integrating over $\omega$ we reduce the r.h.s. of (\ref{label5.28})
to the form
\begin{eqnarray}\label{label5.29}
  \delta {\cal L}_{\rm eff}[\vartheta_{\rm c\ell}(x)] &=& 
  \frac{1}{2}\,\beta^2V[\vartheta_{\rm
    c\ell}(x)]\int^{+\infty}_{-\infty}\frac{dk}{4\pi}
  \frac{1}{\sqrt{\alpha_0 + k^2}} = \nonumber\\
&=&-\,\frac{\beta^2}{4\sqrt{\alpha_0}}\,
V[\vartheta_{\rm
    c\ell}(x)]\int^{+\infty}_{-\infty}\frac{dk}{4\pi}\,\sqrt{\alpha_0 + k^2}
  \,\frac{d\delta(k)}{dk},
\end{eqnarray}
where $\delta(k) $ is a phase shift defined in Eq.(\ref{label5.20}).
We discuss this expression in Section 7 in connection with the
correction to the soliton mass caused by  Gaussian fluctuations.

For the analysis of the renormalizability of the sine--Gordon model
the momentum integral in the effective Lagrangian $\delta {\cal
  L}_{\rm eff}[\vartheta_{\rm c\ell}(x)]$ should be taken in a Lorentz
covariant form (\ref{label5.28}) and regularized in a covariant way.
Making a Wick rotation $\omega \to i\,\omega$ and passing to Euclidean
momentum space we define the integral over $\omega$ and $k$ in
(\ref{label5.28}) as \cite{FI6}
\begin{eqnarray}\label{label5.30}
\int^{+\infty}_{-\infty}\frac{dk}{2\pi}
\int^{+\infty}_{-\infty}\frac{d\omega}{2\pi i }\,\frac{1}{\alpha_0 -
\omega^2 + k^2 - i\,0} = \frac{1}{4\pi}\,{\ell
n}\Big(\frac{\Lambda^2}{\alpha_0}\Big),
\end{eqnarray}
where $\Lambda$ is an Euclidean cut--off \cite{FI6}. The effective
Lagrangian, induced by Gaussian fluctuations around a soliton
solution, is equal to
\begin{eqnarray}\label{label5.31}
\hspace{-0.3in}&&\delta {\cal L}_{\rm eff}[\vartheta_{\rm c\ell}(x)] =
\frac{\alpha_0}{\beta^2}\,\Big[- \frac{\beta^2}{8\pi}\,{\ell
n}\Big(\frac{\Lambda^2}{\alpha_0}\Big)\Big]\,
(\cos\beta\vartheta_{\rm c\ell}(x) - 1).
\end{eqnarray}
The Lagrangian (\ref{label5.31}) has the distinct form of the
correction, caused by quantum fluctuations around the trivial vacuum
calculated to first orders in $\alpha_0(\Lambda^2)$ and $\beta^2$.

The total Lagrangian, accounting for Gaussian  fluctuations
around the soliton  solution amounts to
\begin{eqnarray}\label{label5.32}
{\cal L}_{\rm eff}(x) = \frac{1}{2}\partial_{\mu}\vartheta_{\rm
c\ell}(x)\partial^{\mu}\vartheta_{\rm c\ell}(x) +
\frac{\alpha_0}{\beta^2}\,\Big[1 - \frac{\beta^2}{8\pi}\,{\ell
n}\Big(\frac{\Lambda^2}{\alpha_0}\Big)\Big]\,(\cos\beta\vartheta_{\rm
c\ell}(x) - 1).
\end{eqnarray}
This coincides with Eq.(6.7) of Ref.\cite{FI6}. 

As has been shown in \cite{FI6}, the dependence of the effective
Lagrangian (\ref{label5.32}) on the cut--off $\Lambda$ can be removed
by renormalization with the renormalization constant (\ref{label1.4})
\begin{eqnarray}\label{label5.33}
  \hspace{-0.3in}&&\alpha_0\Big[1 - \frac{\beta^2}{8\pi}\,{\ell
    n}\Big(\frac{\Lambda^2}{\alpha_0}\Big)\Big] = \alpha_r(M^2) Z_1
  \Big[1 - \frac{\beta^2}{8\pi} {\ell
    n}\Big(\frac{\Lambda^2}{\alpha_r(M^2) Z_1}\Big)\Big] = \nonumber\\
  \hspace{-0.3in}&&= \alpha_r(M^2) \Big[1 + 
  \frac{\beta^2}{8\pi}{\ell n}\Big(\frac{\Lambda^2}{M^2}\Big) - 
  \frac{\beta^2}{8\pi}{\ell
    n}\Big(\frac{\Lambda^2}{\alpha_r(M^2)}\Big)\Big] = \alpha_r(M^2)\Big[1 + 
  \frac{\beta^2}{8\pi} {\ell n}\Big(\frac{\alpha_r(M^2)}{M^2}\Big)\Big],
\quad\quad
\end{eqnarray}
where we have kept terms of order $O(\beta^2)$  in the
$\beta^2$--expansion of the renormalization constant (\ref{label1.4}).
This gives the effective Lagrangian
\begin{eqnarray}\label{label5.34}
{\cal L}_{\rm eff}(x) = \frac{1}{2}\partial_{\mu}\vartheta_{\rm
c\ell}(x)\partial^{\mu}\vartheta_{\rm c\ell}(x) +
\frac{\alpha_r(M^2)}{\beta^2}\,\Big[1 + \frac{\beta^2}{8\pi}\,{\ell
n}\Big(\frac{\alpha_r(M^2)}{M^2}\Big)\Big]\,(\cos\beta\vartheta_{\rm
c\ell}(x) - 1).
\end{eqnarray}
We can replace the coupling constant $\alpha_r(M^2)$ by the physical
coupling constant $\alpha_{\rm ph}$, related to $\alpha_r(M^2)$ by
(\ref{label3.19})
\begin{eqnarray}\label{label5.35}
\alpha_{\rm ph} = \alpha_r(M^2)\Big[1 + \frac{\beta^2}{8\pi}\,{\ell
n}\Big(\frac{\alpha_r(M^2)}{M^2}\Big)\Big],
\end{eqnarray}
where we have kept terms of order $O(\beta^2)$ only. Substituting
(\ref{label5.35}) in (\ref{label5.34}) we get
\begin{eqnarray}\label{label5.36}
{\cal L}^{(r)}_{\rm eff}(x) = \frac{1}{2}\partial_{\mu}\vartheta_{\rm
  c\ell}(x)\partial^{\mu}\vartheta_{\rm c\ell}(x) +
\frac{\alpha_{\rm ph}}{\beta^2}\,
(\cos\beta\vartheta_{\rm
  c\ell}(x) - 1).
\end{eqnarray}
The effective Lagrangian (\ref{label5.36}) coincides with the
Lagrangian, renormalized by the quantum fluctuations around the
trivial vacuum (\ref{label3.21}), and corroborates the result obtained
in \cite{FI6}.

We would like to emphasize that analysing the renormalization of the
sine--Gordon model, caused by Gaussian fluctuations around a soliton,
one can see that Gaussian fluctuations are perturbative fluctuations
of order $O(\alpha_r(M^2)\beta^2)$ valid for $\beta^2 \ll 8\pi$, which
cannot be responsible for non--perturbative contributions to the
soliton mass at $\beta^2 = 8\pi$.

\section{Renormalization of the soliton mass by Gaussian fluctuations
  in continuous space--time}
\setcounter{equation}{0}

Using the effective Lagrangian Eq.(\ref{label5.29}) one can calculate
the soliton mass corrected by quantum fluctuations. It
reads 
\begin{eqnarray}\label{label6.1}
 M_s = \frac{8\sqrt{\alpha_0}}{\beta^2} + \Delta M_s,
\end{eqnarray}
where $\Delta M_s$ is defined by
\begin{eqnarray}\label{label6.2}
  \Delta M_s = -\int^{+\infty}_{-\infty}dx^1\,
\delta {\cal L}_{\rm eff}[\vartheta_{\rm c\ell}(x^1)] = 
\int^{+\infty}_{-\infty}\frac{dk}{4\pi}\,\sqrt{\alpha_0 + k^2}
  \,\frac{d\delta(k)}{dk}.
\end{eqnarray}
This corresponds to the correction to the soliton mass, induced by
Gaussian fluctuations, without a {\it surface term}
$-\,\sqrt{\alpha_0}/\pi$ \cite{RR82,AR97}.

In the Lorentz covariant form the correction to the soliton mass reads 
\begin{eqnarray}\label{label6.3}
  &&\Delta M_s = -\int^{+\infty}_{-\infty}dx^1\,
  \delta {\cal L}_{\rm eff}[\vartheta_{\rm c\ell}(x^1)] =\nonumber\\
  &&= \int^{+\infty}_{-\infty}\frac{d\omega}{2\pi i
  }\int^{+\infty}_{-\infty}\frac{dk}{2\pi}\frac{\sqrt{\alpha_0}}{k^2 +
    \alpha_0} [{\ell n}(- \omega^2) - {\ell n}(\alpha_0 - \omega^2 +
  k^2)]\nonumber\\
  &&=- 
  2\sqrt{\alpha_0}\int^{+\infty}_{-\infty}\frac{dk}{2\pi}
  \int^{+\infty}_{-\infty}\frac{d\omega}{2\pi
    i }\frac{1}{\alpha_0 - \omega^2 + k^2 - i\,0},
\end{eqnarray}
where we have taken the effective Lagrangian defined by
(\ref{label5.27}) and integrated over $\omega$ by parts.  Using the
result of the calculation of the integral (\ref{label5.30}) we get the
following expression for the soliton mass corrected by Gaussian
fluctuations
\begin{eqnarray}\label{label6.4}
  M_s = \frac{8\sqrt{\alpha_0(\Lambda^2)}}{\beta^2} - 
\frac{\sqrt{\alpha_0(\Lambda^2)}}{2\pi}\,{\ell
    n}\Big[\frac{\Lambda^2}{\alpha_0(\Lambda^2)}\Big].
\end{eqnarray}
The removal of the dependence on the cut--off $\Lambda$ should be
carried out within the renormalization procedure.

Replacing $\alpha_0(\Lambda^2)$ in (\ref{label6.4}) by
$\alpha_r(M^2) Z_1(\beta^2,M^2; \Lambda^2)$, where the renormalization
constant $Z_1(\beta^2,M^2; \Lambda^2)$ is defined by Eq.(\ref{label1.4}), we get
\begin{eqnarray}\label{label6.5}
  M_s = \frac{8\sqrt{\alpha_r(M^2) Z_1}}{\beta^2} - 
\frac{\sqrt{\alpha_r(M^2)\,Z_1}}{2\pi}\,{\ell
    n}\Big[\frac{\Lambda^2}{\alpha_r(M^2) Z_1}\Big].
\end{eqnarray}
The renormalization constant $Z_1$ should be expanded in power of
$\beta^2$ to order $O(\beta^2)$. This gives
\begin{eqnarray}\label{label6.6}
 Z_1 = 1 + \frac{\beta^2}{8\pi}\,{\ell n}\Big(\frac{\Lambda^2}{M^2}\Big).
\end{eqnarray}
Substituting (\ref{label6.6}) into (\ref{label6.5}) and keeping
only the leading terms in $\beta^2$ we get 
\begin{eqnarray}\label{label6.7}
  M_s = \frac{8}{\beta^2}\,\sqrt{\alpha_r(M^2)\,\Big[1 +
\frac{\beta^2}{8\pi}\,{\ell
    n}\Big(\frac{\alpha_r(M^2)}{M^2}\Big)\Big]}.
\end{eqnarray}
Using Eq.(\ref{label5.35}) we can rewrite the r.h.s. of
(\ref{label6.7}) in terms of $\alpha_{\rm ph}$:
\begin{eqnarray}\label{label6.8}
  M_s = \frac{8\sqrt{\alpha_{\rm ph}}}{\beta^2}.
\end{eqnarray}
The mass of a soliton $M_s$ depends on the physical coupling constant
$\alpha_{\rm ph}$.  Hence, the contribution of Gaussian fluctuations
around a soliton solution is absorbed by the renormalized coupling
constant $\alpha_{\rm ph}$ and no singularities of the sine--Gordon
model appear at $\beta^2 = 8\pi$. 

This result confirms the assertion by Zamolodchikov and Zamolodchikov
\cite{AZ79}, that the singularity of the sine--Gordon model induced by
the finite correction $-\sqrt{\alpha_{\rm ph}}/\pi$ to the soliton
mass, caused by Gaussian fluctuations around a soliton solution, is
completely due to the regularization and renormalization procedure.
This has been  corroborated in \cite{FI6}.

As has been shown in Section 3, a non--trivial finite correction to
the soliton mass   appears due to non--Gaussian quantum fluctuations
of order of $\alpha^2_{\rm ph}\beta^2$ (see Eq.(\ref{label3.31})): 

$$
M_s = \frac{8\sqrt{\alpha_{\rm ph}}}{\beta^2} -
\frac{\sqrt{\alpha_{\rm ph}}}{\pi}. \eqno(3.31) 
$$

\noindent The second term in Eq.(\ref{label3.31}) coincides with that
obtained by Dashen {\it et al.} \cite{RD74,RD75}. However, it is
meaningful only as a perturbative correction for $\beta^2 \ll 8\pi$,
which cannot produce any non--perturbative singularity of the
sine--Gordon model at $\beta^2 = 8\pi$.

We have obtained that the soliton mass $M_s$ does not depend on the
normalization scale $M$.  This testifies that the soliton mass $M_s$
is an observable quantity.

\section{Renormalization of soliton mass by Gaussian fluctuations.
  Space--time discretization technique}
\setcounter{equation}{0}

Usually the correction to the soliton mass is investigated in the
literature by a discretization procedure \cite{RR82} (see also
\cite{AR97}). The soliton with Gaussian fluctuations is embedded into
a spatial box with a finite volume $L$ and various boundary conditions
for Gaussian fluctuations at $x = \pm L/2$. In such a discretization
approach time is also discrete with a period $T$, which finally has to
be taken in the limit $T \to \infty$. The frequency spectrum is
$\omega_m = 2\pi m/T$ with $m = 0, \pm 1,\ldots$. For various boundary
conditions spectra of the momenta of Gaussian fluctuations around a
soliton and of Klein--Gordon quanta, corresponding to vacuum
fluctuations, are adduced in Table 1.
\begin{table}\caption{The spectra of the momenta of Gaussian
    fluctuations around a soliton  and the Klein--Gordon
    quanta. The modes, denoted by $(*)$ are due to the bound state.}

\vspace{0.2in}

\begin{tabular}{llll}
  \hline\hline&&&\\
  {\bf PERIODIC BC:}&&&\\&&&\\
  {\bf Soliton Sector} && {\bf Vacuum Sector}&\phantom{xxxx}\\&&&\\
  $k_nL+\delta(k_n)=2n\pi$ && $q_nL=2n\pi$&\\
  $n=0:(\ast)$ &$\leftarrow1\times\rightarrow$\phantom{xxx}& $n=0:q_0=0$&\\
  $n=1:q_1=\frac{\pi}{L}+\mathcal{O}(L^{-2})$ &$\leftarrow2\times\rightarrow$&
 $n=1:q_1=\frac{2\pi}{L}$&\\
  $\cdots$ &$\leftarrow2\times\rightarrow$& $\cdots$&\\&&&\\
  $\sum_{n=1}^\infty\rightarrow\int_{\frac{\pi}{L}+\mathcal{O}(L^{-2})}^\infty
  \frac{dk}{2\pi}\Bigl(L + 
  \frac{d\delta(k)}{dk}\Bigr)$ &&
  $\sum_{n=1}^\infty\rightarrow\int_{\frac{2\pi}{L}}^\infty\frac{dq}{2\pi}L$&\\&&&\\ 
  \hline\hline&&&\\
  {\bf ANTI-PERIODIC BC:}&&&\\&&&\\ 
  {\bf Soliton Sector} && {\bf Vacuum Sector}&\\&&&\\
  $k_nL+\delta(k_n)= (2 n-1)\pi$ && $q_nL= (2 n-1)\pi$&\\
  $n=1:(\ast) + k_1=0$ &\hspace{-5mm}$\leftarrow(1+1)\times\rightarrow$& 
$n=1:q_1=\frac{\pi}{L}$&\\
  $\cdots$ &$\leftarrow2\times\rightarrow$& $\cdots$&\\&&&\\
  $\sum_{n=2}^\infty\rightarrow\int_{\frac{2\pi}{L}+\mathcal{O}(L^{-2})}^\infty
  \frac{dk}{2\pi}\Bigl(L + 
  \frac{d\delta(k)}{dk}\Bigr)$ &&
  $\sum_{n=2}^\infty\rightarrow\int_{\frac{3\pi}{L}}^\infty\frac{dq}{2\pi}L$&\\&&&\\ 
  \hline\hline&&&\\
  {\bf RIGID WALLS:}&&&\\&&&\\
  {\bf Soliton Sector} && {\bf Vacuum Sector}&\\&&&\\
  $k_nL+\delta(k_n)=n\pi$ && $q_nL=n\pi$&\\
  $n=1:(\ast)$ &$\leftarrow1\times\rightarrow$& $n=1:q_1= \frac{\pi}{L}$&\\
  $\cdots$ &$\leftarrow1\times\rightarrow$& $\cdots$&\\&&&\\
  $\sum_{n=2}^\infty\rightarrow\int_{\frac{\pi}{L}+\mathcal{O}(L^{-2})}^\infty
\frac{dk}{\pi}\Bigl(L + 
  \frac{d\delta(k)}{dk}\Bigr)$ &&
  $\sum_{n=2}^\infty\rightarrow\int_{\frac{2\pi}{L}}^\infty\frac{dq}{\pi}L$&\\&&&\\ 
  \hline\hline
\end{tabular}
\end{table}
According to Table 1, for various boundary conditions the corrections
to the soliton mass are given by
\begin{eqnarray}\label{label7.1}
  \hspace{-0.3in}&&\Delta M^{(\rm p)}_s = \lim_{T \to \infty}\lim_{L \to \infty}\frac{1}{2iT}
  \Big\{2\sum^\infty_{m=-\infty}\sum_{n=1}^\infty\Big[{\ell n}\Big(\alpha_0 - 
  \frac{4\pi^2 m^2}{T^2} + k^2_n\Big) -
  {\ell n}\Big(\alpha_0 - \frac{4\pi^2 m^2}{T^2} + q_n^2\Big)\Big]\nonumber\\
  \hspace{-0.3in}&&+ \sum^\infty_{m=-\infty}
  \Big[{\ell n}\Big( - \frac{4\pi^2 m^2}{T^2} + \Delta^2(L)\Big) - 
  {\ell n}\Big(\alpha_0 - \frac{4\pi^2 m^2}{T^2}\Big)\Big]
  \Big\},\nonumber\\
  \hspace{-0.3in}&&\Delta M^{(\rm ap)}_s = \lim_{T \to \infty}\lim_{L \to \infty}
\frac{1}{2iT}\Big\{2
\sum^\infty_{m=-\infty}\sum_{n=2}^\infty\Big[{\ell n}\Big(\alpha_0 - 
  \frac{4\pi^2 m^2}{T^2} +k^2_n\Big) -
  {\ell n}\Big(\alpha_0 - \frac{4\pi^2 m^2}{T^2} + q_n^2\Big)\Big]\nonumber\\
  \hspace{-0.3in}&&+ \sum^\infty_{m=-\infty}
  \Big[{\ell n}\Big( - \frac{4\pi^2 m^2}{T^2} + \Delta^2(L)\Big) +  
  {\ell n}\Big(\alpha_0 - \frac{4\pi^2 m^2}{T^2}\Big) - 2 {\ell n}
\Big(\alpha_0 - \frac{4\pi^2 m^2}{T^2} + q^2_1\Big)\Big]
  \Big\}, \nonumber\\
  \hspace{-0.3in}&&\Delta M^{(\rm rw)}_s = \lim_{T \to \infty}\lim_{L \to \infty}
\frac{1}{2iT}\Big\{
\sum^\infty_{m=-\infty}\sum_{n=2}^\infty\Big[{\ell n}\Big(\alpha_0 - 
  \frac{4\pi^2 m^2}{T^2} + k^2_n\Big) -
  {\ell n}\Big(\alpha_0 - \frac{4\pi^2 m^2}{T^2} + q_n^2\Big)\Big]\nonumber\\
  \hspace{-0.3in}&&+ \sum^\infty_{m=-\infty}
  \Big[{\ell n}\Big( - \frac{4\pi^2 m^2}{T^2} + \Delta^2(L)\Big) - 
  {\ell n}\Big(\alpha_0 - \frac{4\pi^2 m^2}{T^2} + q^2_1\Big)\Big]
  \Big\},
\end{eqnarray}
where $\Delta^2(L) \sim \alpha_0\,e^{\textstyle -\sqrt{\alpha_0}\,L}$
at $L \to \infty$, and the abbreviations (p), (ap) and (rw) mean  periodic,
anti--periodic boundary conditions and rigid walls, respectively.  

For the summation over $m$ we use the formula, derived by Dolan and
Jackiw \cite{RJ74}:
\begin{eqnarray}\label{label7.2}
  \sum^{+\infty}_{m = -\,\infty}\Big[{\ell n}\Big(\frac{4m^2\pi^2}{-T^2} 
  +  a^2\Big) - {\ell n}\Big(\frac{4m^2\pi^2}{-T^2} +  b^2\Big)\Big] = 
  i\,(a - b)\,T + 2\,{\ell n}\Bigg(\frac{\displaystyle 
    1 - e^{\textstyle\,-i\,a\,T}}{\displaystyle 1 - e^{\textstyle\,-i\,b\,T}}
\Bigg).\quad
\end{eqnarray}
Taking the limit $T \to \infty$ we arrive at the expression 
\begin{eqnarray}\label{label7.3}
 \hspace{-0.03in} \Delta M_s = - \frac{\sqrt{\alpha_0}}{2} +
 \lim_{L \to \infty}\left\{\begin{array}{r@{\quad,\quad}l}
{\displaystyle \sum^{\infty}_{n = 1}
      (\sqrt{\alpha_0 + k^2_n} - 
      \sqrt{\alpha_0 + q^2_n}\,) }& {\rm periodic~BC}\\
{\displaystyle \sum^{\infty}_{n = 2}
      (\sqrt{\alpha_0 + k^2_n} - 
      \sqrt{\alpha_0 + q^2_n}\,) }& {\rm anti-periodic~BC}\\
      {\displaystyle \frac{1}{2}\sum^{\infty}_{n = 2}
      (\sqrt{\alpha_0 + k^2_n} - 
      \sqrt{\alpha_0 + q^2_n}\,)} &{\rm rigid~walls},
\end{array}\right.
\end{eqnarray}
where ${\rm BC}$ is the abbreviation of {\it boundary conditions}.

The aim of our analysis of $\Delta M_s$ within the discretization
procedure is to show that the discretization procedure gives $\Delta
M_s$ in the form of (\ref{label6.3}).

The subsequent analysis of $\Delta M_s$ we carry out for periodic
boundary conditions only. One can show that for anti--periodic
boundary conditions and rigid walls the result is the same.

For the next transformation of the r.h.s. of (\ref{label7.3}) we
propose to use the following integral representation
\begin{eqnarray}\label{label7.4}
  \hspace{-0.05in}\sqrt{\alpha_0 + k^2_n} - 
  \sqrt{\alpha_0 + q^2_n} = \int^{+\infty}_{-\infty}
  \frac{d\omega}{2\pi i}\,
  {\ell n}\Big[\frac{\alpha_0 + k^2_n - \omega^2 -i 0}{\alpha_0 +
 q^2_n - \omega^2 -i 0}\Big].
\end{eqnarray}
This gives
\begin{eqnarray}\label{label7.5}
  \Delta M^{(\rm p)}_s &=& - \frac{\sqrt{\alpha_0}}{2} + 
\lim_{L \to \infty}\sum^{+\infty}_{n = 1}
  \int^{+\infty}_{-\infty}
  \frac{d\omega}{2\pi i}\;
  {\ell n}\Big[\frac{\alpha_0 + k^2_n - \omega^2 -i\,0}{ \alpha_0 
    + q^2_n - \omega^2 -i\,0}\Big] =\nonumber\\
  &=& - \frac{\sqrt{\alpha_0}}{2} + 
  \int^{+\infty}_{-\infty}
  \frac{d\omega}{2\pi i}\lim_{L \to \infty}\sum^{+\infty}_{n = 1}
  {\ell n}\Big[\frac{\alpha_0 + k^2_n - \omega^2 -i\,0}{ \alpha_0 
    + q^2_n - \omega^2 -i\,0}\Big]. 
\end{eqnarray}
To the regularization of the sum over $n$ we apply the {\it
  mode--counting} regularization procedure \cite{AR97}:
\begin{eqnarray}\label{label7.6}
  \hspace{-0.5in}  \Delta M^{(\rm p)}_s = - \frac{\sqrt{\alpha_0}}{2} + 
\int^{+\infty}_{-\infty}
  \frac{d\omega}{2\pi i}\lim_{L \to \infty}\lim_{N\to\infty}\sum^N_{n = 1}
  {\ell n}\Big[\frac{\alpha_0 + k^2_n - \omega^2 -i\,0}{
\alpha_0 + q^2_n - \omega^2 -i\,0}\Big].
\end{eqnarray}
Passing to the continuous momentum representation \cite{AR97} and using
the spectra of the momenta of Gaussian fluctuations and the
Klein--Gordon fluctuations (the vacuum fluctuations) adduced in Table
1 we transcribe the r.h.s of (\ref{label7.6}) into the form
\begin{eqnarray}\label{label7.7}
  \hspace{-0.3in}\Delta M^{(\rm p)}_s &=& - \frac{\sqrt{\alpha_0}}{2} + 
  \int^{+\infty}_{-\infty}
  \frac{d\omega}{2\pi i}\lim_{L \to \infty}\lim_{N\to\infty}
  \Big[\int^{k_N}_{k_1}dk\,\frac{dn(k)}{dk}\,
  {\ell n}(\alpha_0 + k^2 - \omega^2 -i\,0)\nonumber\\  
  \hspace{-0.3in}&&- \int^{q_N}_{q_1}dq\,\frac{dn(q)}{dq}{\ell n}(\alpha_0 + q^2 - \omega^2 -i\,0)\Big] =\nonumber\\  
  \hspace{-0.3in}&&
  - \frac{\sqrt{\alpha_0}}{2} + 
  \int^{+\infty}_{-\infty}
  \frac{d\omega}{2\pi i}\lim_{L \to \infty}\lim_{N\to\infty}
  \Big[\int^{k_N}_{k_1}\frac{dk}{2\pi}\,
  \Big(L + \frac{d\delta(k)}{dk}\Big)\,\nonumber\\
  \hspace{-0.3in}&& \times\,
  {\ell n}(\alpha_0 + k^2 - \omega^2 -i\,0) -  
  L\int^{q_N}_{q_1}\frac{dk}{2\pi}\,{\ell n}(\alpha_0 + k^2 - \omega^2
  -i\,0)\Big], 
\end{eqnarray}
where the limits are equal to (see Table 1):
\begin{eqnarray}\label{label7.8}
 k_1 &=& \frac{\pi}{L}\quad,\quad  k_N = q_N - \frac{\delta(q_N)}{L} =  q_N - 
\frac{\sqrt{\alpha_0}}{\pi N},\nonumber\\
 q_1 &=& \frac{2\pi}{L}\quad,\quad q_N = \frac{2\pi N}{L}.
\end{eqnarray}
Rearranging the limits of integrations we get 
\begin{eqnarray}\label{label7.9}
  \hspace{-0.3in}\Delta M^{(\rm p)}_s &=& - \frac{\sqrt{\alpha_0}}{2} + 
  \int^{+\infty}_{-\infty}
  \frac{d\omega}{2\pi i}\lim_{L \to \infty}
  \lim_{N\to \infty}\int^{k_N}_{\pi/L}\frac{dk}{2\pi}\,
  \frac{d\delta(k)}{dk}\,
  {\ell n}(\alpha_0 + k^2 - \omega^2 -i\,0)\nonumber\\
  \hspace{-0.3in}&&  + \int^{+\infty}_{-\infty}
  \frac{d\omega}{2\pi i}\,\lim_{L \to \infty}L\int^{2\pi/L}_{\pi/L}\frac{dk}{2\pi}
  \,{\ell n}(\alpha_0 + k^2 - \omega^2 -i\,0)\nonumber\\
  \hspace{-0.3in}&& - \int^{+\infty}_{-\infty}
  \frac{d\omega}{2\pi i}\lim_{L \to \infty}\lim_{N\to\infty}L\int^{q_N}_{k_N}
  \frac{dk}{2\pi}\,{\ell n}(\alpha_0 + k^2 - \omega^2 -i\,0) = \nonumber\\
  \hspace{-0.3in}&&= - \frac{\sqrt{\alpha_0}}{2} + \int^{+\infty}_{-\infty}
  \frac{d\omega}{2\pi i}\,\lim_{L \to \infty}\int^{\infty}_{\pi/L}\frac{dk}{2\pi}\,
  \frac{d\delta(k)}{dk}\,
  {\ell n}(\alpha_0 + k^2 - \omega^2 -i\,0)\nonumber\\
  \hspace{-0.3in}&&  + \int^{+\infty}_{-\infty}
  \frac{d\omega}{2\pi i}\,\lim_{L \to \infty}L\int^{2\pi/L}_{\pi/L}\frac{dk}{2\pi}\,
  {\ell n}(\alpha_0 + k^2 - \omega^2 -i\,0)\nonumber\\
  \hspace{-0.3in}&& - \int^{+\infty}_{-\infty}
  \frac{d\omega}{2\pi i}\lim_{L \to \infty}
  \lim_{N\to\infty}L\,\frac{k_N - q_N}{2\pi}\,
  {\ell n}(\alpha_0 + k^{*\,2} - 
  \omega^2 -i\,0).
\end{eqnarray}
For the last term we have applied the {\it mean value theorem} with
$q_N - \delta(q_N)/L < k^* < q_N$.  Since the difference $k_N - q_N =
\delta(q_N)/L = \sqrt{\alpha_0}/\pi N$ is of order $O(1/N)$, this term
vanishes in the limit $N \to \infty$\,\footnote{We would like to
  emphasize that exactly the term of this kind leads to the
  contribution of the finite {\it surface term}
  $-\,\sqrt{\alpha_0}/\pi$ in a regularization procedure using the
  expressions (\ref{label7.3}) without turning to the
  integral representation (\ref{label7.4}).}. As a result we get
\begin{eqnarray}\label{label7.10}
  \hspace{-0.3in}\Delta M^{(\rm p)}_s &=& - \frac{\sqrt{\alpha_0}}{2} + 
\int^{+\infty}_{-\infty}
  \frac{d\omega}{2\pi i}\lim_{L \to \infty}\int^{\infty}_{\pi/L}\frac{dk}{2\pi}\,
\frac{d\delta(k)}{dk}\,
  {\ell n}(\alpha_0 + k^2 - \omega^2 -i\,0)\nonumber\\
  \hspace{-0.3in}&&  + \int^{+\infty}_{-\infty}
  \frac{d\omega}{2\pi i}\,\lim_{L \to \infty}L\int^{2\pi/L}_{\pi/L}\frac{dk}{2\pi}\,
{\ell n}(\alpha_0 + k^2 - \omega^2 -i\,0).
\end{eqnarray}
Taking the limit $L \to \infty$ and applying to the last term the {\it
  mean value theorem} we arrive at the expression
\begin{eqnarray}\label{label7.11}
  \Delta M^{(\rm p)}_s &=& - \frac{\sqrt{\alpha_0}}{2} +  \int^{+\infty}_{-\infty}
  \frac{d\omega}{4\pi i}\,{\ell n}(\alpha_0 - \omega^2 -i\,0)\nonumber\\
  &&+ \int^{+\infty}_{-\infty}
  \frac{d\omega}{2\pi i}\int^{\infty}_0\frac{dk}{2\pi}\,\frac{d\delta(k)}{dk}\,
  {\ell n}(\alpha_0 + k^2 - \omega^2 -i\,0),
\end{eqnarray}
which we transcribe into the form
\begin{eqnarray}\label{label7.12}
  \hspace{-0.3in}&&\Delta M^{(\rm p)}_s = - \frac{\sqrt{\alpha_0}}{2} + \int^{+\infty}_{-\infty}
  \frac{d\omega}{2\pi i}\int^{\infty}_0\frac{dk}{2\pi}\,\frac{d\delta(k)}{dk}\,
  {\ell n}( -\omega^2 -i\,0) +  \int^{+\infty}_{-\infty}
  \frac{d\omega}{4\pi i}\,{\ell n}(\alpha_0 - \omega^2 -i\,0)\nonumber\\
  \hspace{-0.3in}&&+ 
  \int^{+\infty}_{-\infty}
  \frac{d\omega}{2\pi i}\int^{\infty}_0\frac{dk}{2\pi}\,\frac{d\delta(k)}{dk}\,
  [{\ell n}(\alpha_0 + k^2 - \omega^2 -i\,0) - {\ell n}( -\omega^2 -i\,0)].
\end{eqnarray}
The next steps of the reduction of the r.h.s. of (\ref{label7.12}) to
the form (\ref{label6.3}) are rather straightforward. First, one can
easily show that
\begin{eqnarray}\label{label7.13}
  &&\int^{+\infty}_{-\infty}
  \frac{d\omega}{2\pi i}\int^{\infty}_0\frac{dk}{2\pi}\,\frac{d\delta(k)}{dk}\,
  {\ell n}( -\omega^2 -i\,0) +  \int^{+\infty}_{-\infty}
  \frac{d\omega}{4\pi i}\,{\ell n}(\alpha_0 - \omega^2 -i\,0) = \nonumber\\
  &&= \int^{+\infty}_{-\infty}\frac{d\omega}{4\pi i}\,
  [{\ell n}(\alpha_0 - \omega^2 -i\,0) -  {\ell n}( -\omega^2 -i\,0)] =  
  \frac{\sqrt{\alpha_0}}{2}
\end{eqnarray}
and, second, integrating over $\omega$ by parts the last integral in
(\ref{label7.12}) can be reduced to the form
\begin{eqnarray}\label{label7.14}
 \Delta M^{(\rm p)}_s =  - 
  2\sqrt{\alpha_0}\int^{+\infty}_{-\infty}\frac{dk}{2\pi}
  \int^{+\infty}_{-\infty}\frac{d\omega}{2\pi
    i }\frac{1}{\alpha_0 - \omega^2 + k^2 - i\,0}.
\end{eqnarray}
Thus, we have shown that the correction to the soliton mass, induced
by Gaussian fluctuations around a soliton and calculated by means of
the discretization procedure, agrees fully with that we have obtained
in continuous space--time (\ref{label6.3}). 

Hence, the renormalization of the soliton mass, caused by Gaussian
fluctuations calculated within the space--time discretization
technique, coincides with the renormalization of the soliton mass in
continuous space--time. We would like to emphasize that the obtained
result (\ref{label7.14}) does not depend on the boundary conditions.

The calculation of the functional determinant within the
discretization procedure has confirmed the absence of the correction
$-\,\sqrt{\alpha_0}/\pi$. This agrees with our assertion that such a
correction does not appear due to Gaussian fluctuations around a
soliton, corresponding to quantum fluctuations to first orders in
$\alpha_0(\Lambda^2)$ and $\beta^2$. 

The reduction of $\Delta M^{(\rm p)}_s$ of Eq.(\ref{label7.1}) to the expression
(\ref{label6.3}) can be carried out directly. First, summing over $n$
within the {\it mode--counting} regularization procedure and taking
the limit $L\to \infty$ we arrive from $\Delta M^{(\rm p)}_s$ of
Eq.(\ref{label7.1}) at
\begin{eqnarray}\label{label7.15}
  \hspace{-0.3in} &&\Delta M^{(\rm p)}_s = \lim_{T \to \infty}\frac{1}{iT}
  \sum^{\infty}_{m = -\infty}\lim_{L\to \infty}\Big\{\frac{1}{2}\,
  \Big[{\ell n}\Big( -\frac{4\pi^2m^2}{T^2} + \Delta^2(L)\Big) - 
  {\ell n}\Big(\alpha_0  -\frac{4\pi^2m^2}{T^2}\Big)\Big]\nonumber\\
  \hspace{-0.3in} &&+ \lim_{N\to \infty}\sum^N_{n = 1}
  \Big[{\ell n}\Big(\alpha_0  + k^2_n - \frac{4\pi^2m^2}{T^2}\Big) - 
  {\ell n}\Big(\alpha_0 + q^2_n  - \frac{4\pi^2m^2}{T^2}\Big)\Big]\Big\} =
  \nonumber\\
  \hspace{-0.3in} &&= \lim_{T \to \infty}\frac{1}{iT}
  \sum^{\infty}_{m = -\infty}\lim_{L\to \infty}\Big\{\frac{1}{2}
  \Big[{\ell n}\Big( -\frac{4\pi^2m^2}{T^2} + \Delta^2(L)\Big) 
  - {\ell n}\Big(\alpha_0  - \frac{4\pi^2m^2}{T^2}\Big)\Big]\nonumber\\
  \hspace{-0.3in} &&+
  \lim_{N\to \infty}\Big[\int^{k_N}_{k_1}dk\frac{dn(k)}{d k}
  {\ell n}\Big(\alpha_0 + k^2  - \frac{4\pi^2m^2}{T^2}\Big) -
  \int^{q_N}_{q_1}dq \frac{dn(q)}{d q} 
{\ell n}\Big(\alpha_0  + q^2 - \frac{4\pi^2m^2}{T^2}\Big)\Big]\Big\} =\nonumber\\
  \hspace{-0.3in} &&= \lim_{T \to \infty}\frac{1}{iT}
  \sum^{\infty}_{m = -\infty}\lim_{L\to \infty}\Big\{\frac{1}{2}\Big[{\ell n}
  \Big( -\frac{4\pi^2m^2}{T^2} 
  + \Delta^2(L)\Big) - {\ell n}\Big(\alpha_0  - 
  \frac{4\pi^2m^2}{T^2}\Big)\Big]\nonumber\\
  \hspace{-0.3in} &&+ \lim_{N\to \infty}\Big[\int^{k_N}_{\pi/L}\frac{dk}{2\pi}\,
  \frac{d\delta(k)}{d k}\,{\ell n}\Big(\alpha_0 + k^2  - 
  \frac{4\pi^2m^2}{T^2}\Big)+ L\int^{2\pi/L}_{\pi/L}\frac{dk}{2\pi}\,{\ell n}\Big(\alpha_0  + k^2 - \frac{4\pi^2m^2}{T^2}\Big)
  \nonumber\\
  \hspace{-0.3in} && - L\int^{q_N}_{k_N}\frac{dk}{2\pi}\,
  {\ell n}\Big(\alpha_0  - \frac{4\pi^2m^2}{T^2} + k^2\Big)\Big] = \nonumber\\
  \hspace{-0.3in} &&= \lim_{T \to \infty}\frac{1}{iT}
  \sum^{\infty}_{m = -\infty}\lim_{L\to \infty}\Big\{\frac{1}{2}\Big[{\ell n}\Big( -\frac{4\pi^2m^2}{T^2} 
  + \Delta^2(L)\Big) - {\ell n}\Big(\alpha_0  - 
  \frac{4\pi^2m^2}{T^2}\Big)\Big]\nonumber\\
  \hspace{-0.3in} &&+ \int^{\infty}_{\pi/L}\frac{dk}{2\pi}\,
  \frac{d\delta(k)}{d k}\,{\ell n}\Big(\alpha_0 + k^2  -
  \frac{4\pi^2m^2}{T^2}\Big) + L\int^{2\pi/L}_{\pi/L}\frac{dk}{2\pi}\,
{\ell n}\Big(\alpha_0 + k^2  - \frac{4\pi^2m^2}{T^2}\Big)\Big\}=\nonumber\\
  \hspace{-0.3in} &&= \lim_{T \to \infty}\frac{1}{iT}
  \sum^{\infty}_{m = -\infty}\Big\{\frac{1}{2}
\Big[{\ell n}\Big( -\frac{4\pi^2m^2}{T^2}\Big) 
- {\ell n}\Big(\alpha_0  - \frac{4\pi^2m^2}{T^2}\Big)\Big]\nonumber\\
  \hspace{-0.3in} &&+\int^{\infty}_0\frac{dk}{2\pi}\,
  \frac{d\delta(k)}{d k}\,{\ell n}\Big(\alpha_0 + k^2  - 
  \frac{4\pi^2m^2}{T^2}\Big)
  + \frac{1}{2}\,{\ell n}\Big(\alpha_0  - \frac{4\pi^2m^2}{T^2} \Big)\Big\}=
  \nonumber\\
  \hspace{-0.3in} &&= \lim_{T \to \infty}\frac{1}{iT}
  \sum^{\infty}_{m = -\infty}\Big\{\int^{\infty}_0\frac{dk}{2\pi}\,
  \frac{d\delta(k)}{d k}\,{\ell n}\Big(\alpha_0 + k^2  - 
  \frac{4\pi^2m^2}{T^2}\Big) + \frac{1}{2}\,
  {\ell n}\Big(- \frac{4\pi^2m^2}{T^2} \Big)\Big\}.
\end{eqnarray}
Now we  use the integral representation ${^5}$ and get
\begin{eqnarray}\label{label7.16}
  \hspace{-0.3in} &&\Delta M^{(\rm p)}_S = \lim_{T \to \infty}\frac{1}{iT}
  \sum^{\infty}_{m = -\infty}\int^{\infty}_0\frac{dk}{2\pi}\,
\frac{2\sqrt{\alpha_0}}{\alpha_0 + k^2}\,
\Big[{\ell n}\Big(- \frac{4\pi^2m^2}{T^2} \Big) -
  {\ell n}\Big(\alpha_0 + k^2  - 
  \frac{4\pi^2m^2}{T^2}\Big) \Big] =
  \nonumber\\
  \hspace{-0.3in} &&=\lim_{T\to \infty}\int^{+\infty}_{-\infty}\frac{d\omega}{iT}\,
  \frac{dm(\omega)}{d\omega}\int^{\infty}_0\frac{dk}{2\pi}\,\frac{2\sqrt{\alpha_0}}{\alpha_0 + k^2}\,[{\ell n}(- \omega^2) - 
  {\ell n}(\alpha_0 + k^2 - \omega^2)]=\nonumber\\
  \hspace{-0.3in} &&=\int^{+\infty}_{-\infty}\frac{d\omega}{2\pi i}\int^{\infty}_0
  \frac{dk}{2\pi}\,\frac{2\sqrt{\alpha_0}}{\alpha_0 + k^2}\,
  [{\ell n}(- \omega^2) - 
  {\ell n}(\alpha_0 + k^2 - \omega^2)] =\nonumber\\
  \hspace{-0.3in} &&= - 2\sqrt{\alpha_0}
  \int^{+\infty}_{-\infty}\frac{d\omega}{2\pi i}\int^{\infty}_{-\infty}
  \frac{dk}{2\pi}\,\frac{1}{\alpha_0 - \omega^2 + k^2 -i\,0}.
\end{eqnarray}
For the other boundary conditions we get the same result.

Thus, we have shown that the discretized version of the correction to
the soliton mass reduces to the continuum result if one transcribes
first the sum over the quantum number $n$ of the momenta of Gaussian
and vacuum fluctuations into the corresponding integral over the
momentum $k$.

\section{Conclusion}

We have investigated the renormalizability of the sine--Gordon model.
We have analysed the renormalizability of the two--point Green
function to second order in $\alpha$ and to all orders in $\beta^2$.
We have shown that the divergences appearing in the sine--Gordon model
can be removed by the renormalization of the dimensional coupling
constant $\alpha_0(\Lambda^2)$.  We remind that the coupling constant
$\beta^2$ is not renormalizable. This agrees well with a possible
interpretation of the coupling constant $\beta^2$ as $\hbar$
\cite{SC75,FI1}. The quantum fluctuations calculated to first order in
$\alpha_r(M^2)$, the renormalized dimensional coupling constant
depending on the normalization scale $M$, and to arbitrary order in
$\beta^2$ after removal of divergences form a physical coupling
constant $\alpha_{\rm ph}$, which is finite and does not depend on the
normalization scale $M$. We have argued that the total renormalized
two--point Green function depends on the physical coupling constant
$\alpha_{\rm ph}$ only. In order to illustrate this assertion (i) we
have calculated the correction to the two--point Green function to
second order in $\alpha_r(M^2)$ and to all orders in $\beta^2$ and
(ii) we have solved the Callan--Symanzik equation for the two--point
Green function with the Gell--Mann--Low function, defined to all
orders in $\alpha_r(M^2)$ and $\beta^2$. We have found that the
two--point Green function of the sine--Gordon field depends on the
running coupling constant $\alpha_r(p^2) = \alpha_{\rm
  ph}(p^2/\alpha_{\rm ph})^{\tilde{\beta}^2/8\pi}$, where
$\tilde{\beta}^2 = \beta^2/(1 + \beta^2/8\pi) < 1$ for any $\beta^2$.

We have shown that the finite contribution of the quantum
fluctuations, calculated to second order in $\alpha_{\rm ph}$ and to
first order in $\beta^2$, leads to an effective Lagrangian

$$
{\cal L}(x) = \frac{1}{2}\,\partial_{\mu}\vartheta(x)
\partial^{\mu}\vartheta(x) + \frac{\alpha_{\rm eff}}{\beta^2}\,
(\cos\beta\vartheta(x) - 1)\eqno(3.30)
 $$

\noindent with the effective coupling constant $\alpha_{\rm eff} =
\alpha_{\rm ph}\,(1 - \beta^2/4\pi)$, which defines the soliton  mass

$$
  M_s = \frac{8\sqrt{\alpha_{\rm eff}}}{\beta^2} = 
\frac{8\sqrt{\alpha_{\rm ph}}}{\beta^2} - 
\frac{\sqrt{\alpha_{\rm ph}}}{\pi}.\eqno(3.31)
$$

\noindent The term $-\,\sqrt{\alpha_{\rm ph}}/\pi$ is a perturbative
finite correction to the soliton mass, which coincides with the result
obtained by Dashen {\it et al.} \cite{RD74,RD75}. Such a correction
has been interpreted in the literature as a non--perturbative
contribution leading to a singularity of the sine--Gordon model at
$\beta^2 = 8\pi$ (see also \cite{LD78}).  However, in our approach
such a correction is a perturbative one valid for $\beta^2 \ll 8\pi$
and does not provide any singularity for the sine--Gordon model in the
non--perturbative regime $\beta^2 \sim 8\pi$. This confirms the
conjecture by Zamolodchikov and Zamolodchikov \cite{AZ79} that a
singularity of the sine--Gordon model at $\beta^2 = 8\pi$
\cite{RD74,RD75,LD78} is a superficial one and depends on the
regularization and renormalization procedure. This has been
corroborated in \cite{FI6}.

In addition to the analysis of the renormalizability of the
sine--Gordon model with respect to quantum fluctuations relative to
the trivial vacuum, we have analysed the renormalizability of the
sine--Gordon model with respect to quantum fluctuations around a
soliton.  Following Dashen {\it et al.}  \cite{RD74,RD75} and Faddeev
and Korepin \cite{LD78} we have taken into account only Gaussian
fluctuations. 

For the calculation of the effective Lagrangian, induced by Gaussian
fluctuations, we have used the path--integral approach and integrated
over the field $\varphi(x)$, fluctuating around a soliton.  This has
allowed to express the effective Lagrangian in terms of the functional
determinant.  For the calculation of the contribution of the
functional determinant we have used the eigenfunctions and eigenvalues
of the differential operator, describing the evolution of the field
$\varphi(x)$. We have shown that the renormalized effective
Lagrangian, induced by Gaussian fluctuations around a soliton,
coincides completely with the renormalized Lagrangian of the
sine--Gordon model, caused by quantum fluctuations around the trivial
vacuum to first order in $\alpha_0$ and to second order in $\beta^2$.
After the removal of divergences the total effective Lagrangian,
caused by Gaussian fluctuations around a soliton, is equal to
(\ref{label3.21})

$$
{\cal L}(x) = \frac{1}{2}\,\partial_{\mu}\vartheta(x)
\partial^{\mu}\vartheta(x) +
\frac{\alpha_{\rm ph}}{\beta^2}\,
(\cos\beta\vartheta(x) - 1).\eqno(3.21)
$$

\noindent This implies that Gaussian fluctuations around a soliton do
not produce any quantum corrections to the soliton mass.  After the
removal divergences the soliton mass is equal to the mass of a
soliton, calculated without quantum corrections, with the replacement
$\alpha_0 \to \alpha_{\rm ph}$.  Hence, no non--perturbative
singularities of the sine Gordon model at $\beta^2 = 8\pi$ can be
induced by Gaussian fluctuations around a soliton.

For the confirmation of our results, obtained in continuous
space--time, we have calculated the functional determinant caused by
Gaussian fluctuations around a soliton within the discretization
procedure with periodic and anti--periodic boundary conditions and
rigid walls. We have shown that the result of the calculation of the
functional determinant (i) coincides with that we have obtained in
continuous space--time and (ii) does not depend on the boundary
conditions.

\newpage

\section*{Appendix. Solutions of the differential equation for
  Gaussian fluctuations around a soliton }

Making a change of variables $\xi = \tanh(\sqrt{\alpha_0}\sigma)$ we
reduce (\ref{label5.15}) to the form \cite{LL65}

$$
\frac{d}{d\xi}\Big[(1 - \xi^2)\frac{d\psi(\xi)}{d \xi}\Big] +\Big[s(s
+ 1) - \frac{\epsilon^2}{1 - \xi^2}\Big]\psi(\xi) = 0,\eqno({ \rm
A}.1)
$$

\noindent where $s(s + 1) = 2$ and $\epsilon^2 = - k^2/\alpha_0 = 1 -
\omega^2/\alpha_0$ \cite{LL65}.

Substituting $\psi(\xi) = (1 - \xi^2)^{\epsilon/2}\,w(\xi)$ and
denoting $u = (1 - \xi)/2$ we arrive at the equation \cite{LL65}

$$
u(1 - u)w\,'' + (\epsilon + 1)(1 - 2u)\,w\,' - (\epsilon - s)(\epsilon
+ s + 1)\,w = 0.\eqno({ \rm A}.2)
$$

\noindent The solution of this equation can be given in terms of hypergeometric
functions $F(a,b; c; z)$ \cite{NL73} (see also \cite{HMF72}). It reads

$$
w(\xi) = w^{(1)}(\xi) + w^{(2)}(\xi) = C_1\,F\Big(\epsilon - s, \epsilon + s
+ 1; \epsilon + 1;\frac{1 - \xi}{2}\Big)
$$
$$
+ C_2\,\Big(\frac{1 - \xi}{2}\Big)^{-\epsilon} \,F\Big( - s, s + 1; 1
- \epsilon; \frac{1 - \xi}{2}\Big),\eqno({ \rm A}.3)
$$

\noindent where $C_1$ and $C_2$ are the integration constants.

The parameter $s$ acquires two values $s = -2, + 1$, which are
solutions of the equation $s(s + 1) = 2$. Since for both case the
hypergeometric functions coincide \cite{NL73}, setting $s = 1$ we
obtain

$$
w(\xi) = w^{(1)}(\xi) + w^{(2)}(\xi) = C_1\,
F\Big(\epsilon - 1, \epsilon + 2; \epsilon + 1;\frac{1 - \xi}{2}\Big)
$$
$$
+ C_2\,\Big(\frac{1 - \xi}{2}\Big)^{-\epsilon} \,F\Big( - 1, 2;
1 - \epsilon; \frac{1 - \xi}{2}\Big).\eqno({ \rm A}.4) 
$$

\noindent For arbitrary $\epsilon$ the solution $w^{(2)}(\xi)$ can be
reduced to the polynomial \cite{NL73}

$$
w^{(2)}(\xi) = C_2\,\Big(\frac{1 -
  \xi}{2}\Big)^{-\epsilon}\,F\Big( 2, -1; 1 - \epsilon; \frac{1 -
  \xi}{2}\Big) = 
$$
$$
= C_2\,\Big(\frac{1 -
  \xi}{2}\Big)^{-\epsilon}\,F\Big( - 1, 2; 1 - \epsilon; \frac{1 -
  \xi}{2}\Big) = C_2\,\Big(\frac{1 -
  \xi}{2}\Big)^{-\epsilon}\,\frac{\xi - \epsilon}{1 -
  \epsilon}.\eqno({ \rm A}.5) 
$$

\noindent One can show that the solution $w^{(1)}(\xi)$ can be reduced
to a polynomial too. For this aim we suggest to use the following
relation for hypergeometric functions \cite{HMF72,NL73}

$$
F\Big(a,b;c;\frac{1 - \xi}{2}\Big) =
\frac{\Gamma(c)\Gamma(c - a - b)}{\Gamma(c - a) \Gamma(c -
  b)}\,F\Big(a,b;a + b - c + 1; \frac{1 + \xi}{2}\Big) 
$$
$$
+
\Big(\frac{1 + \xi}{2}\Big)^{c - a - b}\,\frac{\Gamma(c)\Gamma(a + b -
  c)}{\Gamma(a) \Gamma( b)}\,F\Big(c - a, c - b; c - a - b + 1;
\frac{1 + \xi}{2}\Big).\eqno({ \rm A}.6) 
$$

\noindent For the solution
$w^{(1)}(\xi)$ we get 

$$
w^{(1)}(\xi) = C_1\,F\Big(\epsilon - 1,
\epsilon + 2; \epsilon + 1;\frac{1 - \xi}{2}\Big) =
C_1\,\frac{\Gamma(\epsilon + 1)\Gamma(-\,\epsilon)}{\Gamma(-1)
}\,F\Big(\epsilon - 1, \epsilon + 2; \epsilon + 1; \frac{1 +
  \xi}{2}\Big) 
$$
$$
+ C_1\,\frac{\Gamma(\epsilon +
  1)\Gamma(\epsilon)}{\Gamma(\epsilon - 1) \Gamma(\epsilon + 2)}\,\Big(\frac{1 +
  \xi}{2}\Big)^{-\epsilon}\,F\Big(2, - 1; 1 - \epsilon; \frac{1 +
  \xi}{2}\Big) = C_1\,\Big(\frac{1 +
  \xi}{2}\Big)^{-\epsilon}\,\frac{\epsilon + \xi}{1 +
  \epsilon}.\eqno({ \rm A}.7) 
$$

\noindent Thus, the fluctuating field
$\psi(\xi)$ is equal to 

$$
\psi(\xi) =
C_1\,2^{\,\epsilon}\,\Big(\frac{1 - \xi}{1 +
  \xi}\Big)^{\epsilon/2}\,\frac{\epsilon + \xi}{1 + \epsilon} +
C_2\,2^{\,\epsilon}\,\Big(\frac{1 + \xi}{1 -
  \xi}\Big)^{\epsilon/2}\,\frac{ \xi - \epsilon }{1 -
  \epsilon}.\eqno({ \rm A}.8) 
$$

\noindent In terms of $\sigma$ the classical
solution for the fluctuating field reads 

$$
\psi(\sigma) =
C_1\,(\epsilon + \tanh(\sqrt{\alpha_0}\sigma))\,e^{\textstyle\,-
  \epsilon \sqrt{\alpha_0} \sigma} + C_2\,(- \epsilon +
\tanh(\sqrt{\alpha_0}\sigma))\,e^{\textstyle\,+ \epsilon
  \sqrt{\alpha_0}\sigma},\eqno({ \rm A}.9) 
$$

\noindent where we have redefined
the integration constants.  These solutions describe a bound state for
$\epsilon = \pm 1$ and a scattering state for $\epsilon = \pm
ik/\sqrt{\alpha_0}$. They are

$$
\psi_b(\sigma) = \sqrt{\frac{\sqrt{\alpha_0}}{2}}\,
\frac{1}{\cosh(\sqrt{\alpha_0}\sigma)}, 
$$
$$
\psi_k(\sigma) =
\frac{i}{\sqrt{2\pi}}\,\frac{-ik +
  \sqrt{\alpha_0}\,\tanh(\sqrt{\alpha_0}\sigma)}{\sqrt{k^2 +
    \alpha_0}}\,e^{\textstyle\,+ ik\sigma}.\eqno({ \rm A}.10) 
$$

\noindent The wave functions $\psi_k(\sigma)$, given by ({\rm A}.10),
are normalized to the $\delta$--function as 

$$
\int^{+\infty}_{-\infty}d\sigma\, \psi^*_{k\,'}(\sigma)
\psi_{k}(\sigma) = \frac{1}{k\,'\,^2 -
  k^2}\,\Big(\psi^*_{k\,'}(\sigma) \frac{d}{d\sigma}\psi_{k}(\sigma) -
\psi_{k}(\sigma)\frac{d}{d \sigma}
\psi^*_{k\,'}(\sigma)\Big)\Big|^{+\infty}_{-\infty} =\delta(k\,' -
k).\eqno({ \rm A}.11) 
$$

\noindent The fluctuating field $\varphi(\tau,\sigma)$
is equal to 

$$
\varphi_{\omega b}(\tau,\sigma) =
\frac{1}{\sqrt{2\pi}}\,\sqrt{\frac{\sqrt{\alpha_0}}{2}}\,
\frac{1}{\cosh(\sqrt{\alpha_0}\sigma)}\, e^{\textstyle\,-i\omega \tau
}, 
$$
$$
\varphi_{\omega k}(\tau,\sigma) = \frac{i}{2\pi}\,\frac{-ik +
  \sqrt{\alpha_0}\,\tanh(\sqrt{\alpha_0}\sigma)}{\sqrt{k^2 +
    \alpha_0}}\,e^{\textstyle\,-i\omega \tau + ik\sigma}. \eqno({ \rm
  A}.12) 
$$

\noindent They are normalized by 

$$
\int^{+\infty}_{-\infty}\int^{+\infty}_{-\infty}d\tau\varphi^*_{\omega\,'
  b}(\tau,\sigma)\varphi_{\omega b}(\tau,\sigma) = \delta(\omega\,' -
\omega), 
$$
$$
\int^{+\infty}_{-\infty}\int^{+\infty}_{-\infty}d\tau
d\sigma\, \varphi^*_{\omega\,' k\,'}(\tau,\sigma) \varphi_{\omega
  k}(\tau,\sigma) = \delta(\omega\,' - \omega)\delta(k\,' - k).\eqno({
  \rm A}.13) 
$$

\noindent The eigenfunction $\psi_b(\sigma)$ has the eigenvalue zero,
$\omega = 0$. This corresponds to the bound state \cite{JR70}. The
eigenfunction $\psi_b(\sigma)$ is normalized to unity

$$
\int^{+\infty}_{-\infty}d\sigma\,|\psi_b(\sigma)|^2 = 1.\eqno({ \rm
  A}.14) 
$$

\noindent The solutions ({\rm A}.10) satisfy the completeness
condition \cite{JR70} 

$$
\int^{+\infty}_{-\infty}dk\,\psi^*_k(\sigma\,'\,)\psi_k(\sigma) +
\psi_b(\sigma\,'\,)\psi_b(\sigma) = \delta(\sigma\,' - \sigma).
\eqno({ \rm A}.15) 
$$

\noindent The proof of the completeness condition ({\rm
  A}.15) 

$$
\int^{+\infty}_{-\infty}dk\,\psi^*_k(\sigma\,'\,)\psi_k(\sigma) +
\psi_b(\sigma\,'\,)\psi_b(\sigma) =
\int^{+\infty}_{-\infty}\frac{dk}{2\pi}\,e^{\textstyle ik(\sigma -
  \sigma\,')} 
$$
$$
+ \sqrt{\alpha_0}\,[\tanh(\sqrt{\alpha_0}\sigma) -
\tanh(\sqrt{\alpha_0}\sigma\,'\,)]\int^{+\infty}_{-\infty}
\frac{dk}{2\pi}\,\frac{ik}{k^2
  + \alpha_0}\,e^{\textstyle ik(\sigma - \sigma\,')} 
$$
$$
+
\alpha_0\, [\tanh(\sqrt{\alpha_0}\sigma\,'\,)
\tanh(\sqrt{\alpha_0}\sigma) - 1]
\int^{+\infty}_{-\infty}\frac{dk}{2\pi}\,\frac{1}{k^2 +
  \alpha_0}\,e^{\textstyle ik(\sigma - \sigma\,')} 
$$
$$
+
\frac{\sqrt{\alpha_0}}{2}\,\frac{1}{\cosh(\sqrt{\alpha_0}\sigma\,'\,)}\,\frac{1}{\cosh(\sqrt{\alpha_0}\sigma)}.
\eqno({\rm A}.16) 
$$

\noindent Integrating over $k$ we get 

$$
\int^{+\infty}_{-\infty}dk\,\psi^*_k(\sigma\,'\,)\psi_k(\sigma) +
\psi_b(\sigma\,'\,)\psi_b(\sigma) = \delta(\sigma\,' - \sigma) -
\frac{\sqrt{\alpha_0}}{2}\,\frac{1}{\cosh(\sqrt{\alpha_0}\sigma\,'\,)}\,\frac{1}{\cosh(\sqrt{\alpha_0}\sigma)}
$$
$$
\times\,e^{\textstyle\,-\sqrt{\alpha_0}|\sigma -
  \sigma\,'\,|}\,[\varepsilon(\sigma -
\sigma\,'\,)\,\sinh(\sqrt{\alpha_0}(\sigma - \sigma\,'\,)) +
\cosh(\sqrt{\alpha_0}(\sigma - \sigma\,'\,))] 
$$
$$
+
\frac{\sqrt{\alpha_0}}{2}\,\frac{1}{\cosh(\sqrt{\alpha_0}
\sigma\,'\,)}\,\frac{1}{\cosh(\sqrt{\alpha_0}\sigma)}
= \delta(\sigma\,' - \sigma),\eqno({ \rm A}.17) 
$$

\noindent where $\varepsilon(\sigma - \sigma\,'\,)$ is the
sign--function. The term, proportional to
$e^{\textstyle\,-\sqrt{\alpha_0}|\sigma - \sigma\,'\,|}$, is given by
the contributions of the second and the third terms of the r.h.s. of
({\rm A}.16). For the derivation of ({\rm A}.17) we have used the
relation

$$
\varepsilon(\sigma - \sigma\,'\,)\,\sinh(\sqrt{\alpha_0}((\sigma -
\sigma\,'\,)) + \cosh(\sqrt{\alpha_0}(\sigma - \sigma\,'\,)) =
e^{\textstyle\,+ \sqrt{\alpha_0}|\sigma - \sigma\,'\,|}.\eqno({ \rm
  A}.17)
 $$

\noindent Thus, the contribution of the second and the third terms
in the r.h.s. of ({\rm A}.16) cancel the contribution of the
zero--mode $\psi_b(\sigma)$. This completes the proof of the
completeness condition ({\rm A}.15).

\end{document}